\documentclass[onecolumn]{IEEEtran}
\usepackage[mathscr]{eucal}
\usepackage[cmex10]{amsmath}
\usepackage{epsfig,epsf,psfrag}
\usepackage{amssymb,amsmath,amsthm,amsfonts,latexsym}
\usepackage{amsmath,graphicx,bm,xcolor,url,overpic}
\usepackage{fixltx2e}
\usepackage{array}
\usepackage{verbatim}
\usepackage{bm}
\usepackage{algorithmic}
\usepackage{algorithm}
\usepackage{verbatim}
\usepackage{textcomp}
\usepackage{mathrsfs}
\usepackage{epstopdf}

\newcommand{\openone}{\leavevmode\hbox{\small1\normalsize\kern-.33em1}}

\catcode`~=11 \def\UrlSpecials{\do\~{\kern -.15em\lower .7ex\hbox{~}\kern .04em}} \catcode`~=13 

\allowdisplaybreaks[4]

\newcommand{\nn}{\nonumber}

\newcommand{\calA}{\mathcal{A}}
\newcommand{\calB}{\mathcal{B}}

\newcommand{\calJ}{\mathcal{J}}
\newcommand{\calK}{\mathcal{K}}
\newcommand{\calL}{\mathcal{L}}
\newcommand{\calM}{\mathcal{M}}

\newcommand{\calP}{\mathcal{P}}

\newcommand{\calT}{\mathcal{T}}

\newcommand{\calX}{\mathcal{X}}
\newcommand{\calY}{\mathcal{Y}}

\newcommand{\bP}{\mathbf{P}}

\newcommand{\bQ}{\mathbf{Q}}

\newcommand{\bx}{\mathbf{x}}
\newcommand{\bX}{\mathbf{X}}

\newcommand{\rmc}{\mathrm{c}}

\newcommand{\rmd}{\mathrm{d}}

\newcommand{\rmG}{\mathrm{G}}

\newcommand{\rmH}{\mathrm{H}}

\newcommand{\rmQ}{\mathrm{Q}}
\newcommand{\rmr}{\mathrm{r}}

\newcommand{\rmT}{\mathrm{T}}

\newcommand{\rmV}{\mathrm{V}}

\newcommand{\bbN}{\mathbb{N}}

\newcommand{\bbP}{\mathbb{P}}

\newcommand{\bbR}{\mathbb{R}}

\DeclareMathAlphabet{\mathbsf}{OT1}{cmss}{bx}{n}
\DeclareMathAlphabet{\mathssf}{OT1}{cmss}{m}{sl}

\DeclareSymbolFont{bsfletters}{OT1}{cmss}{bx}{n}  
\DeclareSymbolFont{ssfletters}{OT1}{cmss}{m}{n}
\DeclareMathSymbol{\bsfGamma}{0}{bsfletters}{'000}
\DeclareMathSymbol{\ssfGamma}{0}{ssfletters}{'000}
\DeclareMathSymbol{\bsfDelta}{0}{bsfletters}{'001}
\DeclareMathSymbol{\ssfDelta}{0}{ssfletters}{'001}
\DeclareMathSymbol{\bsfTheta}{0}{bsfletters}{'002}
\DeclareMathSymbol{\ssfTheta}{0}{ssfletters}{'002}
\DeclareMathSymbol{\bsfLambda}{0}{bsfletters}{'003}
\DeclareMathSymbol{\ssfLambda}{0}{ssfletters}{'003}
\DeclareMathSymbol{\bsfXi}{0}{bsfletters}{'004}
\DeclareMathSymbol{\ssfXi}{0}{ssfletters}{'004}
\DeclareMathSymbol{\bsfPi}{0}{bsfletters}{'005}
\DeclareMathSymbol{\ssfPi}{0}{ssfletters}{'005}
\DeclareMathSymbol{\bsfSigma}{0}{bsfletters}{'006}
\DeclareMathSymbol{\ssfSigma}{0}{ssfletters}{'006}
\DeclareMathSymbol{\bsfUpsilon}{0}{bsfletters}{'007}
\DeclareMathSymbol{\ssfUpsilon}{0}{ssfletters}{'007}
\DeclareMathSymbol{\bsfPhi}{0}{bsfletters}{'010}
\DeclareMathSymbol{\ssfPhi}{0}{ssfletters}{'010}
\DeclareMathSymbol{\bsfPsi}{0}{bsfletters}{'011}
\DeclareMathSymbol{\ssfPsi}{0}{ssfletters}{'011}
\DeclareMathSymbol{\bsfOmega}{0}{bsfletters}{'012}
\DeclareMathSymbol{\ssfOmega}{0}{ssfletters}{'012}

\newcommand{\tilh}{\tilde{h}}

\newcommand{\tilP}{\tilde{P}}

\newcommand{\tilQ}{\tilde{Q}}

\newcommand{\hatT}{\hat{T}}

\newcommand{\barQ}{\bar{Q}}

\DeclareMathOperator*{\argmin}{arg\,min}

\DeclareMathOperator{\supp}{supp}

\newtheorem{theorem}{Theorem} 
\newtheorem{lemma}[theorem]{Lemma}

\newtheorem{proposition}[theorem]{Proposition}
\newtheorem{corollary}[theorem]{Corollary}

\usepackage{subfigure,epstopdf,bbm} 
 
\begin{document}

\title{Second-Order Asymptotically Optimal Statistical Classification}
\author{\IEEEauthorblockN{Lin Zhou, Vincent Y.~F.~Tan and Mehul Motani} 
\thanks{The authors are with the Department of Electrical and Computer Engineering, National University of Singapore (Emails: lzhou@u.nus.edu, vtan@nus.edu.sg, motani@nus.edu.sg). Vincent Y.~F.~Tan is also with the Department of Mathematics, National University of Singapore.}
}

\maketitle

\begin{abstract}
Motivated by real-world machine learning applications,  we analyze approximations to the non-asymptotic fundamental limits of statistical classification. In the binary version of this problem, given two training sequences  generated according to two {\em unknown} distributions $P_1$ and $P_2$, one is tasked to classify a test sequence which is known to be generated according to either $P_1$ or $P_2$. This problem can be thought of as an analogue of the binary hypothesis testing problem but in the present setting, the generating distributions are unknown. Due to  finite sample considerations, we consider the second-order asymptotics (or dispersion-type) tradeoff between type-I and type-II error probabilities for tests which ensure that (i) the type-I error probability for {\em all} pairs of distributions decays exponentially fast and (ii) the type-II error probability for a {\em particular} pair of distributions is non-vanishing. 
We generalize our results to classification of multiple hypotheses with the rejection option.
\end{abstract}


\section{Introduction}

In the   simple binary hypothesis testing problem, one is given a source sequence $Y^n$ and one knows that it is either generated in an i.i.d.\ fashion from one of two known distributions $P_1$ or $P_2$. One is then asked to design a test to make this decision. There is a natural trade-off between the type-I and type-II error probabilities. This is quantified by the Chernoff-Stein lemma~\cite{chernoff1952measure}   in the Neyman-Pearson setting in which the type-I error probability decays exponentially fast in $n$ with exponent given by $D(P_2\| P_1)$ if the type-II error probability is upper bounded by some fixed $\varepsilon \in (0,1)$. Blahut~\cite{blahut1974hypothesis} established the tradeoff between the exponents of the type-I and type-II error probabilities. Strassen~\cite{strassen1962asymptotische} derived a refinement  of the Chernoff-Stein lemma. This area of study is now commonly known as   {\em second-order asymptotics} and it  quantifies the backoff from $D(P_2\|P_1)$ one incurs at finite sample sizes and non-vanishing type-II error probabilities $\varepsilon\in (0,1)$. In all these analyses, the likelihood ratio test~\cite{poor2013introduction} is optimal. 

However, in real-world machine learning applications, the generating distributions are \emph{not} known. For the {\em binary} classification framework, one is given two training sequences, one generated from $P_1$ and the other from $P_2$. Using these training sequences, one attempts to classify a test sequence according to whether one believes that it is generated from either $P_1$ or $P_2$.

\vspace{-.01in}
\subsection{Main Contributions} \vspace{-.01in}
Instead of algorithms, in this paper, we are concerned with the  information-theoretic  limits of the binary classification problem. This was first considered by Gutman who proposed a type-based (empirical distribution-based) test~\cite[Eq.~(6)]{gutman1989asymptotically} and proved that this test is asymptotically optimal in the sense that any other test that achieves the same exponential decay for the type-I error probability for {\em all} pairs of distributions, necessarily has a larger type-II error probability for any {\em fixed} pair of distributions. Inspired by Gutman's~\cite{gutman1989asymptotically} and Strassen's~\cite{strassen1962asymptotische} seminal works, and by practical applications where the number of training and test samples is limited (due to the prohibitive cost in obtaining labeled data),  we derive refinements to the tradeoff between the type-I and type-II error probabilities for such tests. In particular, we derive the exact second-order asymptotics~\cite{strassen1962asymptotische,polyanskiy2010finite,hayashi2009information} for binary classification. Our main result asserts that Gutman's test is second-order optimal. The proofs follow by judiciously modifying  and refining Gutman's arguments in~\cite{gutman1989asymptotically} in both the achievability and converse proofs. In the achievability part, we apply a Taylor expansion to a generalized form of the Jensen-Shannon divergence~\cite{lin1991divergence} and apply the Berry-Esseen theorem to analyze Gutman's test. The converse part follows by showing that Gutman's type-based test is approximately optimal in a certain sense to be made precise in Lemma~\ref{anytotype}.   This study provides intuition for the non-asymptotic fundamental limits and our results have the potential to  allow practitioners to gauge the effectiveness of various classification algorithms. 


Second, we discuss three consequences  of our main result. The first asserts that the largest exponential decay rate of the maximal type-I error probability is a generalized version of the Jensen-Shannon divergence, defined in~\eqref{def:GJS} to follow. This result can be seen as a counterpart of Chernoff-Stein lemma~\cite{chernoff1952measure} which is  applicable   to binary hypothesis testing. Next, we show that our main result can be applied to obtain a second-order asymptotic expansion for the fundamental limits of the two sample homogeneity testing problem~\cite[Sec.~II-C]{unnikrishnan2016weak} and the closeness testing problem~\cite{batu2013testing,acharya2014sublinear,chan2014optimal}. Finally, we consider the dual setting of the main result in which the type-I error probabilities are non-vanishing while the type-II error probabilities decay exponentially fast. In this case, the largest exponential decay rate of the type-II error probabilities for Gutman's rule is given by a  R\'enyi divergence~\cite{renyi1961measures} of a certain order related to the ratio of the lengths of the training and test sequences.

Finally, we generalize our second-order asymptotic result for binary classification to classification of multiple hypotheses with the rejection option. We first consider tests satisfying the following conditions (i) the error probability under each hypothesis decays exponentially fast with the same exponent {\em for all tuples of distributions} and (ii) the rejection probability under each hypothesis is upper bounded by a different constant {\em for a particular tuple}. We derive second-order  approximations of the largest error exponent  for all hypotheses and show that a generalization of Gutman's test by Unnikrishnan in \cite[Theorem~4.1]{unnikrishnan2015asymptotically} is second-order optimal. The proofs follow by generalizing those for binary classification and carefully analyzing the rejection probabilities. In addition, similarly to the binary case, we also consider a dual setting, in which under each hypothesis, the error probability is non-vanishing for all tuples of distributions and the rejection probability decays exponentially fast for a particular tuple. 

\subsection{Related Works}

The most related work is \cite{gutman1989asymptotically} where Gutman showed that his   type-based test is asymptotically optimal for the binary classification problem and its extension to classification of multiple hypotheses with rejection for Markov sources. Ziv~\cite{ziv1988classification} illustrated the relationship between binary classification and universal data compression. The Bayesian setting of the binary classification problem was studied by Merhav and Ziv~\cite{merhav1991bayesian}. Subsequently, Kelly, Wagner, Tularak and Viswanath~\cite{kelly2013} considered the binary classification problem with large alphabets.  Unnikrishnan~\cite{unnikrishnan2015asymptotically} generalized the result of Gutman by considering classification for multiple hypotheses where there are multiple test sequences. Finally, Unnikrishnan and Huang~\cite{unnikrishnan2016weak} approximated the type-I error probability of the binary classification problem using   weak convergence analysis.

\subsection{Organization of the Rest of the Paper}
The rest of our paper is organized as follows. In Section \ref{sec:formulation}, we set up the notation, formulate the binary classification 
problem and present existing results by Gutman \cite{gutman1989asymptotically}. In Section \ref{sec:results4bc}, we discuss the motivation for our setting and present our second-order    result for binary classification. We also discuss some consequences of our main result. In Section \ref{sec:result4cmr}, we generalize our result for binary classification to classification of multiple hypotheses with the rejection option. The proofs of our results are provided in Section \ref{sec:proofs}. The proofs of some supporting lemmas are deferred to the appendices.

\section{Problem Formulation and Existing Results}
\label{sec:formulation}
\subsection{Notation} \label{sec:notation}
Random variables and their realizations are in upper (e.g.,  $X$) and lower case (e.g.,  $x$) respectively. All sets are denoted in calligraphic font (e.g.,  $\mathcal{X}$). We use $\calX^{\mathrm{c}}$ to denote the complement of $\calX$. Let $X^n:=(X_1,\ldots,X_n)$ be a random vector of length $n$. All logarithms are base $e$. We use $\Phi(\cdot)$ to denote the  cumulative distribution function (cdf) of the standard Gaussian and  $\Phi^{-1}(\cdot )$ its inverse. Let $\rmQ(t):=1-\Phi(t)$ be the corresponding complementary cdf.  We use $\rmG_k(\cdot)$ to denote the complementary cdf of a chi-squared random variable with $k$ degrees of freedom and $\rmG^{-1}_k(\cdot)$ its inverse. Given any two integers $(a,b)\in\bbN^2$, we use $[a:b]$ to denote the set of integers $\{a,a+1,\ldots,b\}$ and use $[a]$ to denote $[1:a]$. The set of all probability distributions on a finite set $\calX$ is denoted as $\calP(\calX)$. Notation concerning the method of types follows~\cite{Tanbook}. Given a vector $x^n = (x_1,x_2,\ldots,x_n) \in\calX^n$, the {\em type} or {\em empirical distribution} is denoted as $\hat{T}_{x^n}(a)=\frac{1}{n}\sum_{i=1}^n \mathbbm{1}\{x_i=a\},a\in\calX$. The set of types formed from length-$n$ sequences with alphabet $\calX$ is denoted as $\calP_{n}(\calX)$. Given $P\in\calP_{n}(\calX)$, the set of all sequences of length $n$ with type $P$, the {\em type class}, is denoted as $\calT^n_P$. The {\em support} of the probability mass function $P\in\calP(\calX)$ is denoted as $\supp(P):=\{x\in\calX:P(x)>0\}$. 

\subsection{Problem Formulation}
The main goal in  {\em binary hypothesis testing}   is to classify a sequence $Y^n$ as being independently generated from one of   two distinct distributions $(P_1,P_2)\in\calP(\calX)^2$. However, different from classical binary hypothesis testing~\cite{lehmann2006testing,blahut1974hypothesis} where the two distributions are known, in {\em binary classification}~\cite{gutman1989asymptotically}, we do not know the two distributions. We instead have two training sequences $X_1^N$ and $X_2^N$ generated  in an i.i.d.\ fashion according to $P_1$ and $P_2$ respectively. Therefore, the two hypotheses are 
\begin{itemize}
\item $\rmH_1$: the test sequence $Y^n$ and the 1$^{\mathrm{st}}$ training sequence $X_1^N$ are generated according to the same distribution;
\item $\rmH_2$: the test sequence $Y^n$ and the 2$^{\mathrm{nd}}$ training sequence  $X_2^N$ are generated according to the same distribution.
\end{itemize}
We assume that $N=\lceil\alpha n\rceil$ for some $\alpha\in\bbR_+$.\footnote{\label{foot} In the following, we will often write $N=n\alpha$ for brevity, ignoring the integer constraints on $N$ and $n$.} The task in the binary classification problem is to design a decision rule (test) $\phi_n:\calX^{2N}\times\calX^n\to\{\rmH_1,\rmH_2\}$. Note that a decision rule partitions the sample space $\calX^{2N}\times\calX^n$ into two disjoint regions:  $\calA(\phi_n)$ where any triple $(X_1^N,X_2^N,Y^n)\in\calA(\phi_n)$ favors hypothesis $\rmH_1$ and  $\calA^\rmc(\phi_n)$ where any triple $(X_1^N,X_2^N,Y^n)\in\calA^\rmc(\phi_n)$ favors hypothesis $\rmH_2$. 

Given any decision rule $\phi_n$ and any pair of distributions $(P_1,P_2)\in\calP(\calX)^2$, we have two types of error probabilities, i.e.,
\begin{align}
\beta_1(\phi_n|P_1,P_2)
&:=\bbP_1\big\{\phi_n(X_1^N,X_2^N,Y^n)=\rmH_2\big\}
\label{def:type1err},\\
\beta_2(\phi_n|P_1,P_2)
&:=\bbP_2\big\{\phi_n(X_1^N,X_2^N,Y^n)=\rmH_1\big\}
\label{def:type2err},
\end{align}
where for $j\in[2]$, we define $\bbP_j\{\cdot\}:=\Pr\{\cdot|\rmH_j\}$ where $X_i^N\sim P_i^N$ for all $i\in[2]$. The two error probabilities in \eqref{def:type1err} and \eqref{def:type2err} are respectively known as the type-I and type-II error probabilities.

\subsection{Existing Results and Definitions}
The goal of binary classification is to design a classification  rule based on the training sequences. This rule is then used on the test sequence to decide whether $\rmH_1$ or $\rmH_2$ is true. We revisit the study of the fundamental limits of the problem here. Towards this goal, Gutman~\cite{gutman1989asymptotically} proposed a decision rule using marginal types of $X_1^N$, $X_2^N$ and $Y^n$. To present Gutman's test, we need the following generalization of the Jensen-Shannon divergence~\cite{lin1991divergence}. Given any two distributions $(P_1,P_2)\in\calP(\calX)^2$ and any number $\alpha\in\bbR_+$, let the {\em generalized Jensen-Shannon divergence} be
\begin{align}
\mathrm{GJS}(P_1,P_2,\alpha)
&:=\alpha D\Big(P_1\Big\|\frac{\alpha P_1+P_2}{1+\alpha}\Big)+D\Big(P_2\Big\|\frac{\alpha P_1+P_2}{1+\alpha}\Big)\label{def:GJS}.
\end{align}
Given a threshold $\lambda\in\bbR_+$ and any triple $(x_1^N,x_2^N,y^n)$,  Gutman's decision rule is as follows:
\begin{align}
\phi_n^{\rm{Gut}}(x_1^N,x_2^N,y^n)
&:=\left\{
\begin{array}{ll}
\rmH_1&\mathrm{if}~\mathrm{GJS}(\hatT_{x_1^N},\hatT_{y^n},\alpha)\leq \lambda\\
\rmH_2&\mathrm{if}~\mathrm{GJS}(\hatT_{x_1^N},\hatT_{y^n},\alpha)>\lambda.
\end{array}
\right.\label{gutmanrule}
\end{align}
To state Gutman's main result, we define the following ``exponent'' function
\begin{align}
F(P_1,P_2,\alpha,\lambda)
&:=\min_{\substack{(Q_1,Q_2)\in\calP(\calX)^2:\\\mathrm{GJS}(Q_1,Q_2,\alpha)\leq \lambda}} \alpha D(Q_1\|P_1)+ D(Q_2\|P_2)\label{def:FP1P2l}.
\end{align}
Note that $F(P_1,P_2,\alpha,\lambda)=0$ for $\lambda\geq \mathrm{GJS}(P_1,P_2,\alpha)$ and that $\lambda \mapsto F(P_1,P_2,\alpha,\lambda)$ is continuous (a consequence of \cite[Lemma~12]{Tan11_IT} in which $y\mapsto \min_{x\in \calK}f(x,y)$ is continuous if $f$ is continuous and $\calK$ is compact).

Gutman~\cite[Lemma 2 and Theorem 1]{gutman1989asymptotically} showed that the   rule in \eqref{gutmanrule} is asymptotically optimal (error exponent-wise) if the type-I error probability vanishes exponentially fast over {\em all pairs of distributions}.

\begin{theorem}
\label{gutmantheorem}
Gutman's decision rule $\phi_n^{\rm{Gut}}$  satisfies the following two properties:
\begin{enumerate} 
\item \emph{Asymptotic/Exponential performance}: For any pair of distributions $(P_1,P_2)\in\calP(\calX)^2$,
\begin{align}
\liminf_{n\to\infty} -\frac{1}{n}\log \beta_1(\phi_n^{\rm{Gut}}|P_1,P_2)&\geq \lambda,\label{gutet1}\\*
\liminf_{n\to\infty} -\frac{1}{n}\log \beta_2(\phi_n^{\rm{Gut}}|P_1,P_2)&\geq F(P_1,P_2,\alpha,\lambda).\label{gutet2}
\end{align}
\item  \emph{Asymptotic/Exponential Optimality}: Fix a sequence of decision rules  $\{\phi_n\}_{n=1}^\infty$ such that for all pairs of distributions $(\tilP_1,\tilP_2)\in\calP(\calX)^2$,
\begin{align}
\liminf_{n\to\infty}-\frac{1}{n}\log\beta_1(\phi_n|\tilP_1,\tilP_2)\ge \lambda,
\end{align}
then for any pair of distributions $(P_1,P_2)\in\calP(\calX)^2$,
\begin{align}
\beta_2(\phi_n|P_1,P_2)\geq \beta_2(\phi_n^{\rm{Gut}}|P_1,P_2),
\end{align}
where $\phi_n^{\rm{Gut}}$ is Gutman's test with threshold $\lambda$ defined in \eqref{gutmanrule} which achieves~\eqref{gutet1}--\eqref{gutet2}. 
\end{enumerate}
\end{theorem}
We remark that using Sanov's theorem~\cite[Chapter 11]{cover2012elements}, one can easily show that, for any pairs of distributions $(P_1,P_2)\in\calP(\calX)^2$ and any $\lambda>0$, Gutman's decision rule in \eqref{gutmanrule} satisfies \eqref{gutet1} as well as 
\begin{equation}
\lim_{n\to\infty} -\frac{1}{n}\log \beta_2(\phi_n^{\rm{Gut}}|P_1,P_2)=F(P_1,P_2,\alpha,\lambda).
\end{equation}
Note that   Theorem~\ref{gutmantheorem} is analogous to  Blahut's work~\cite{blahut1974hypothesis} in which the trade-off of the error exponents for the binary hypothesis testing problem was thoroughly analyzed. 


\section{Binary Classification}
\label{sec:results4bc}

\subsection{Definitions and Motivation} \label{sec:def}
In this paper, motivated by practical applications where the lengths of source sequences are finite (obtaining labeled training samples is prohibitively expensive), we are interested in approximating the non-asymptotic fundamental limits in terms of the tradeoff between type-I and type-II error probabilities of optimal tests. In particular,   out of all tests whose type-I error probabilities decay exponentially fast for all pairs of distributions and whose type-II error probability is upper bounded by a constant $\varepsilon\in(0,1)$ for a particular pair of distributions, what is the largest decay rate of the sequence of the type-I error probabilities? In other words, we are interested in the following fundamental limit
\begin{align}
\lambda^*(n,\alpha,\varepsilon|P_1,P_2)
\nn:=\sup\Big\{\lambda\in\bbR_+:\exists~\phi_n~\mathrm{s.t.~}\beta_1(\phi_n|\tilP_1,\tilP_2)&\leq \exp(-n\lambda),~\forall~(\tilP_1,\tilP_2)\in\calP(\calX)^2,\\
\mathrm{and~}\beta_2(\phi_n|P_1,P_2)&\leq \varepsilon\Big\}\label{def:l1^*}.
\end{align}
From Theorem \ref{gutmantheorem} (see also \cite[Theorem 3]{gutman1989asymptotically}), we obtain that 
\begin{align}
\liminf_{n\to\infty}\lambda^*(n,\alpha,\varepsilon|P_1,P_2)\geq \mathrm{GJS}(P_1,P_2,\alpha)\label{firstordergutman}.
\end{align}
As a corollary of our result in Theorem \ref{bc:second},  we find that the result in \eqref{firstordergutman} is in fact tight and the limit exists.
  In this paper, we refine the above asymptotic statement and, in particular, provide second-order approximations to $\lambda^*(n,\alpha,\varepsilon|P_1,P_2)$.

To conclude this section, we explain why we consider $\lambda^*(n,\alpha,\varepsilon|P_1,P_2)$ instead of characterizing a seemingly more natural quantity, namely, the largest decay rate of type-I error probability when the type-II error probability is upper bounded by a constant $\varepsilon\in(0,1)$ for a particular pair of distributions $(P_1,P_2)\in\calP(\calX)^2$, i.e.,
\begin{align}
\beta_2^*(n,\alpha,\varepsilon|P_1,P_2)
&:=\inf\Big\{r\in[0,1]:\exists~\phi_n~\mathrm{s.t.~}\beta_1(\phi_n|P_1,P_2)\leq r,~\beta_2(\phi_n|P_1,P_2)\leq \varepsilon\Big\}.
\end{align}
In the binary classification problem, when we design a test $\phi_n$, we do not know the pair of distributions $(P_1,P_2)\in\calP(\calX)^2$ from which the training sequences are generated. Thus, unlike the simple hypothesis testing problem~\cite{strassen1962asymptotische,csiszar2011information}, we cannot design of a test tailored to a particular pair of distributions. Instead, we are interested in designing {\em universal} tests which have good performances {\em for all} pairs of distributions for the type-I (resp.\ type-II) error probability  and at the same time, constrain the type-II (resp.\ type-I) error probability  with respect to a {\em particular} pair of distributions $(P_1,P_2)$. 

\subsection{Main Result}\label{sec:main_res}
We need the following definitions before presenting our main result. Given any $x\in\calX$ and any pair of distributions $(P_1,P_2)\in\calP(\calX)^2$, define the following two {\em information densities}
\begin{align}
\imath_i(x|P_1,P_2,\alpha)&:=\log\frac{(1+\alpha)P_i(x)}{\alpha P_1(x)+P_2(x)},\quad i\in[2]\label{def:i}.
\end{align}
Furthermore, given any pair of distributions $(P_1,P_2)\in\calP(\calX)^2$, define the following {\em dispersion function} (linear combination of the variances of the information densities)
\begin{align}
\rmV(P_1,P_2,\alpha)
&=\alpha\mathrm{Var}_{P_1}[\imath_1(X|P_1,P_2,\alpha)]+\mathrm{Var}_{P_2}[\imath_2(X|P_1,P_2,\alpha)]\label{def:v}.
\end{align}

\begin{theorem}
\label{bc:second}
For any $\varepsilon\in(0,1)$, any $\alpha\in\bbR_+$ and any pair of distributions $(P_1,P_2)\in\calP(\calX)^2$, we have
\begin{align}
\lambda^*(n,\alpha,\varepsilon|P_1,P_2) =\mathrm{GJS}(P_1,P_2,\alpha)+\sqrt{\frac{\rmV(P_1,P_2,\alpha)}{n}}\Phi^{-1}(\varepsilon)+O\left(\frac{\log n}{n}\right). \label{eqn:lambda_s}
\end{align}
\end{theorem}
Theorem \ref{bc:second} is proved in Section \ref{proof:bcsecond}. 
In~\eqref{eqn:lambda_s}, $\mathrm{GJS}(P_1,P_2,\alpha)$ and $\sqrt{{\rmV(P_1,P_2,\alpha)}/{n}}\,\Phi^{-1}(\varepsilon)$ are respectively known as the {\em first-} and {\em second-order terms} in the {\em asymptotic expansion} of   $\lambda^*(n,\alpha,\varepsilon|P_1,P_2)$. Since $0<\varepsilon<1/2$ in most applications, $\Phi^{-1}(\varepsilon)<0$ and so the second-order term represents a {\em backoff} from the exponent $\mathrm{GJS}(P_1,P_2,\alpha)$ at finite sample sizes $n$.  As shown by Polyanskiy, Poor and Verd\'u~\cite{polyanskiy2010finite} (also see~\cite{polyanskiy2010thesis}),  in the channel coding context, these two terms usually constitute a reasonable approximation to the non-asymptotic fundamental limit  at moderate $n$.  This will also be corroborated  numerically for the current problem in Section~\ref{sec:num}. Several other remarks are in order.

First, we remark that since the achievability part is based on Gutman's test, this test in~\eqref{gutmanrule} is second-order optimal. This means that it achieves the optimal second-order term in the asymptotic expansion of $\lambda^*(n,\alpha,\varepsilon|P_1,P_2)$. 

Second, as a corollary of our result, we obtain that  for any $\varepsilon\in(0,1)$, 
\begin{align}
\lim_{n\to\infty} \lambda^*(n,\alpha,\varepsilon|P_1,P_2)=\mathrm{GJS}(P_1,P_2,\alpha).
\end{align}
In other words, a {\em strong converse} for $\lambda^*(n,\alpha,\varepsilon|P_1,P_2)$ holds. This result can be understood as the counterpart of the Chernoff-Stein lemma~\cite{chernoff1952measure} for the binary classification problem (with strong converse). In the following, we comment on the influence of the ratio of the number of training and test samples $\alpha = N/n$ in terms of the dominant term in $\lambda^*(n,\alpha,\varepsilon|P_1,P_2)$. Note that the generalized Jensen-Shannon divergence $\mathrm{GJS}(P_1,P_2,\alpha)$   admits the following properties: 
\begin{itemize}
\item[(i)] $\mathrm{GJS}(P_1,P_2,\alpha)$ is increasing in $\alpha$;
\item[(ii)] $\mathrm{GJS}(P_1,P_2,0)=0$ and $\lim_{\alpha\to\infty}\mathrm{GJS}(P_1,P_2,\alpha)=D(P_2\|P_1)$. 
\end{itemize}
Thus, we conclude that the longer the lengths of training sequences (relative to the test sequence), the better the performance in terms of exponential decay rate of type-I error probabilities for all pairs of distributions. In the extreme case in which $\alpha\to 0$, i.e., the training sequence is arbitrarily short compared to the test sequence, we conclude that type-I error probability cannot decay exponentially fast. However, in the other extreme in which  $\alpha\to\infty$, we conclude that type-I error probabilities for all pairs of distributions decay exponentially fast with the dominant (first-order) term being $D(P_2\|P_1)$. This implies that we can achieve the optimal decay rate determined by the Chernoff-Stein lemma~\cite{chernoff1952measure} for binary hypothesis testing. Intuitively, this occurs since when $\alpha\to\infty$, we can estimate the true pair of distributions with arbitrarily high accuracy (using the large number training samples). In fact, we can say even more. Based on the formula in \eqref{def:v}, we deduce that, $\lim_{\alpha \to\infty} \rmV(P_1,P_2,\alpha)= \mathrm{Var}_{P_2} [\log(P_2(X)/P_1(X))]$, the {\em relative entropy variance}, so we recover Strassen's seminal result~\cite[Theorem~1.1]{strassen1962asymptotische} concerning the second-order asymptotics of binary hypothesis testing. 

Finally, we remark that the binary classification problem is closely related with the so-called two sample homogeneity testing problem~\cite[Sec.~II-C]{unnikrishnan2016weak} and the closeness testing problem~\cite{batu2013testing,acharya2014sublinear,chan2014optimal} where given two i.i.d.\ generated sequences $X^N$ and $Y^n$, one aims to determine whether the two sequences are generated according to the same distribution or not. Thus, in this problem, we have the following two hypotheses:
\begin{itemize}
\item $\rmH_1$: the two sequences $X^N$ and $Y^n$ are generated according to the same distribution;
\item $\rmH_2$: the two sequences $X^N$ and $Y^n$ are generated according to different distributions.
\end{itemize}
The task in such a problem is to design a test $\phi_n:\calX^N\times\calY^n\to\rm\{H_1,H_2\}$. Given any   $\phi_n$ and any   $(P_1,P_2)\in\calP(\calX)^2$, the false-alarm and miss detection probabilities for such a problem  are
\begin{align}
\beta_{\rm{FA}}(\phi_n|P_1)
&:=\bbP_{P_1}\big\{\phi_n(X^N,Y^n)=\rmH_2\big\},\\
\beta_{\rm{MD}}(\phi_n|P_1,P_2)
&:=\bbP_{P_1,P_2}\big\{\phi_n(X^N,Y^n)=\rmH_1\big\},
\end{align}
where in $\bbP_{P_1}\{\cdot\}$, the random variables $X^N$ and $Y^n$ are both distributed i.i.d.\ according to $P_1$ and in $\bbP_{P_1,P_2}\{\cdot\}$, $X^N$ and $Y^n$ are distributed i.i.d. according to $P_1$ and $P_2$ respectively. Paralleling our setting for the binary classification problem, we can study the following fundamental limit of the two sample hypothesis testing problem:
\begin{align}
\xi^*(n,\alpha,\varepsilon|P_1,P_2)
\nn:=\sup\Big\{\lambda\in\bbR_+:
\exists~\phi_n\mathrm{~s.t.~}\beta_{\rm{FA}}(\phi_n|\tilP_1)&\leq \exp(-n\lambda),\forall\,\tilP_1\in\calP(\calX),\\*
\beta_{\rm{MD}}(\phi_n|P_1,P_2)&\leq \varepsilon
\Big\}.\label{tsht}
\end{align}
\begin{corollary}
\label{cor:testing}
For any $\varepsilon\in(0,1)$, any $\alpha\in\bbR_+$ and any  $(P_1,P_2)\in\calP(\calX)^2$, we have
\begin{align}
\xi^*(n,\alpha,\varepsilon|P_1,P_2)&=\mathrm{GJS}(P_1,P_2,\alpha)+\sqrt{\frac{\rmV(P_1,P_2,\alpha)}{n}}\Phi^{-1}(\varepsilon)+O\left(\frac{\log n}{n}\right).
\end{align}
\end{corollary}
Since the proof is similar to that of Theorem \ref{bc:second}, we omit it.
 Corollary \ref{cor:testing} implies that Gutman's test is second-order optimal for the two sample homogeneity testing problem. We remark that for the binary classification problem without rejection (i.e., we are not allowed to declare the neither $\rmH_1$ nor $\rmH_2$ is true), the problem is essentially the same as the two sample hypothesis testing problem except that we have one more training sequence. However, as shown in Theorem \ref{bc:second}, the second training sequence is not useful in order to obtain second-order optimal result. This asymmetry in binary classification problem is circumvented if one also considers a rejection option as will be demonstrated in Section \ref{sec:result4cmr}.

\begin{figure}[t]
\centering
\begin{tabular}{cc}
\hspace{-.25in} \includegraphics[width=.5\columnwidth]{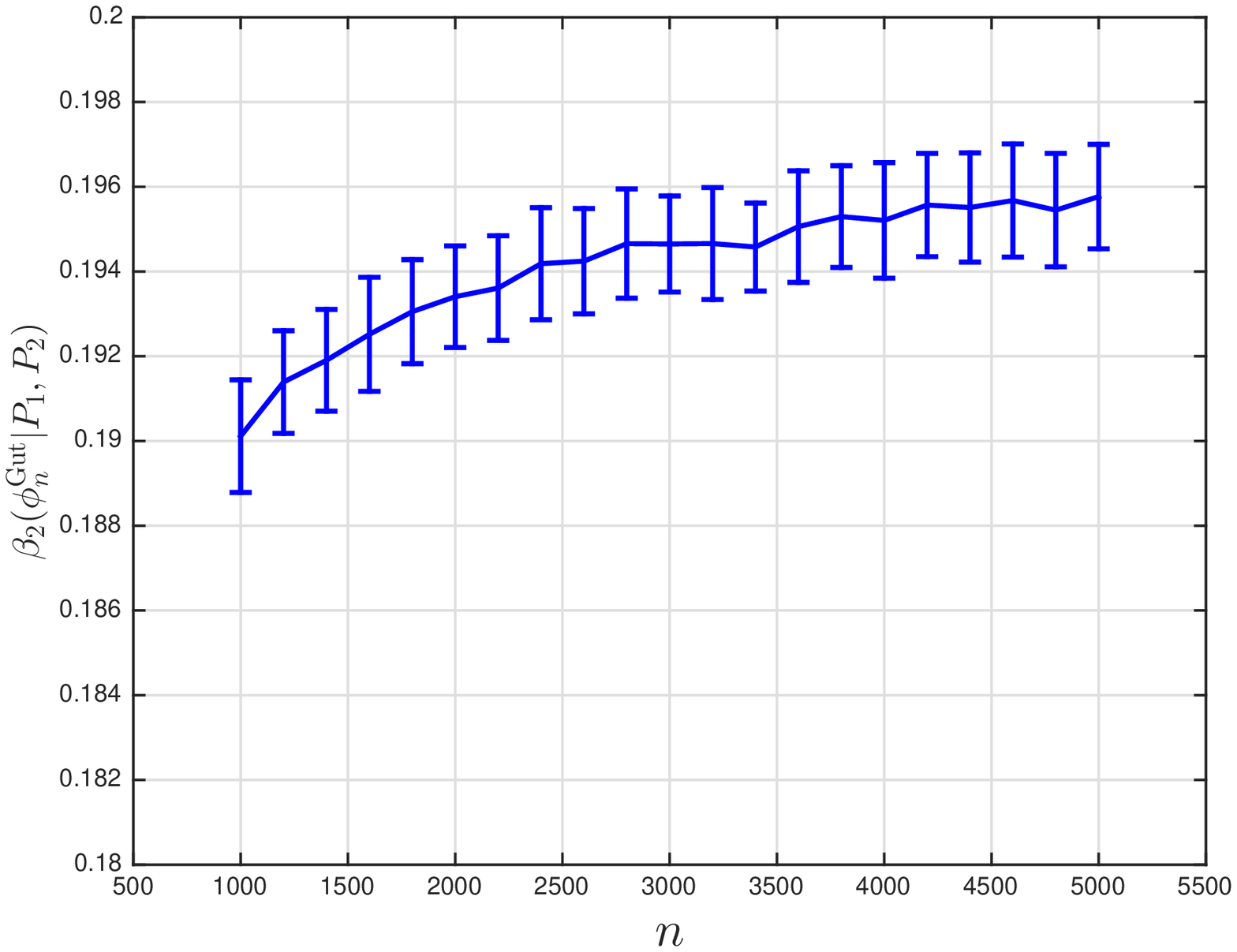}& \hspace{-.4in} \includegraphics[width=.5\columnwidth]{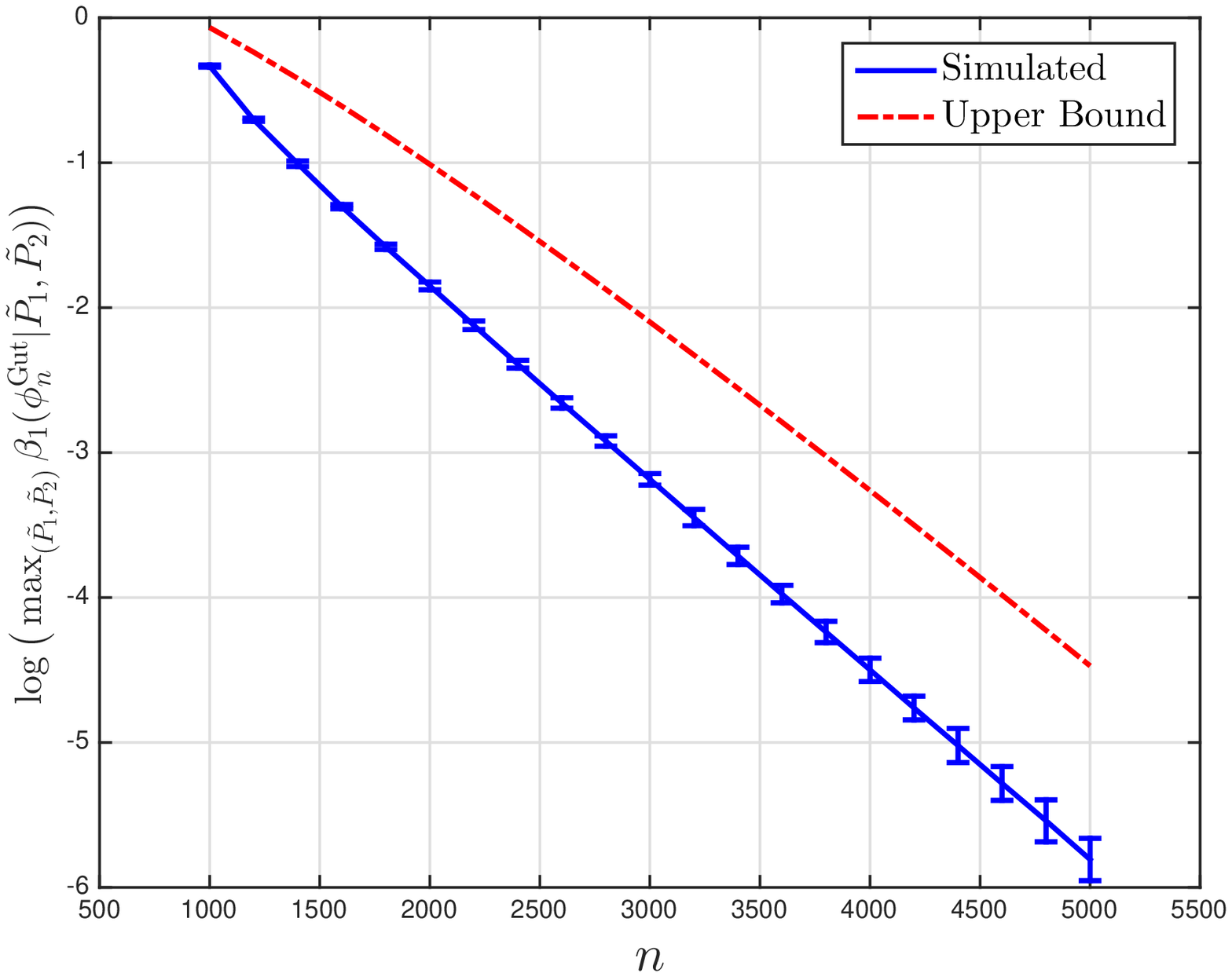}\\
\hspace{-.25in} {\footnotesize (a)  Type-II Error Probability} & \hspace{-.4in}  {\footnotesize (b) Logarithm of the  Maximal Type-I Error Probability}
\end{tabular}
\caption{(a) Type-II error probability for Gutman's test with target error probability $\varepsilon=0.2$. The error bars denote $1$ standard deviation above and below the mean over the   independent experiments; (b) Natural logarithm of the maximal type-I error probability for Gutman's test. The error bars denote $10$ standard deviations above and below the mean. }
 \label{type14Gutman}
\end{figure}

\subsection{Numerical Simulation for Theorem \ref{bc:second}} \label{sec:num}
In this subsection, we present a numerical example to illustrate the performance of Gutman's test in~\eqref{gutmanrule} and the accuracy of our theoretical results. We consider binary sources with alphabet $\calX=\{0,1\}$. Throughout this subsection, we set $\alpha=2$.

 In Figure~\ref{type14Gutman}(a), we plot the type-II error probability $\beta_2(\phi_n^{\rm{Gut}}|P_1,P_2)$ for a particular pair of distributions $(P_1,P_2)$ where $P_1=\mathrm{Bern}(0.2)$ and $P_2=\mathrm{Bern}(0.4)$. The threshold is chosen to be the second-order asymptotic expansion 
\begin{equation}
\hat{\lambda}:=\mathrm{GJS}(P_1,P_2,\alpha)+\sqrt{\frac{\rmV(P_1,P_2,\alpha)}{n}}\Phi^{-1}(\varepsilon),\label{eqn:lambda_hat}
\end{equation}
with target error probability being set to $\varepsilon=0.2$. Each point in Figure~\ref{type14Gutman}(a) is obtained by estimating  the average error probability in the following manner. For each length of the test sequence $n\in \{1000,1200, 1400,\ldots, 5000\}$, we  estimate the type-II error probability of  a single Gutman's test  in~\eqref{gutmanrule} using $10^7$ independent experiments.  From Figure~\ref{type14Gutman}(a), we observe that the simulated error probability for Gutman's test is close to the target error probability  of $\varepsilon=0.2$ as the length of the test sequence $n$ increases. We believe that there is a slight bias in the results as we have not taken the third-order term, which scales as $O(\frac{\log n}{n})$ into account in the threshold in~\eqref{eqn:lambda_hat}.

In Figure \ref{type14Gutman}(b), we plot the natural logarithm of the theoretical upper bound $\exp(-n \hat{\lambda})$ and the maximal empirical type-I error probability $\beta_1(\phi_n^{\rm{Gut}}|\tilP_1,\tilP_2)$ over all pairs of distributions $(\tilP_1,\tilP_2)$. We set the fixed pair of distributions $(P_1,P_2)$ to be $P_1=\mathrm{Bern}(0.2)$ and $P_2=\mathrm{Bern}(0.228)$ and choose $\varepsilon=0.2$. We ensured that the threshold $\hat{\lambda}$ in~\eqref{eqn:lambda_hat} is small enough so that even if $n$ is large, the type-I error event occurs sufficiently many times and thus the numerical results are statistically significant. From Figure~\ref{type14Gutman}(b), we observe that the simulated probability lies below the theoretical one as expected. The gap can be explained by the fact that the method of types analysis is typically loose non-asymptotically due to a large polynomial factor. A more refined analysis based on strong large deviations~\cite[Theorem~3.7.2]{dembo2009large}  would yield better estimates on exponentially decaying probabilities but we do not pursue this here. However, we do note that as $n$ becomes large, the slopes of the simulated and theoretical curves become increasingly close to each other (simulated slope at $n=5000$ is $\approx-0.001336$; theoretical slope at $n=5000$ is  $\approx-0.001225$), showing that on the exponential scale, our estimate of the maximal type-I error probability is relatively tight.

\subsection{Analysis of Gutman's Test in A Dual Setting}
\label{sec:weakconvergence}

In addition to analyzing $\lambda^*(n,\alpha,\varepsilon|P_1,P_2)$, one might also be interested in decision rules whose type-I error probabilities for all pairs of distributions are non-vanishing and whose type-II error probabilities for a particular pair of distributions decays exponentially fast. To be specific, for any decision rule $\phi_n$, we consider the following non-asymptotic fundamental limit:
\begin{align}
\tau^*(n,\alpha,\varepsilon|\phi_n,P_1,P_2)
\nn&:=\sup\Big\{\tau\in\bbR_+: \beta_1(\phi_n|\tilP_1,\tilP_2)\leq \varepsilon,~\forall~(\tilP_1,\tilP_2)\in\calP(\calX)^2\\*
&\qquad\quad\qquad\qquad\quad\mathrm{and~}\beta_2(\phi_n|P_1,P_2)\leq \exp(-n\tau)~\Big\}\label{def:l2^*}.
\end{align}
This can be considered as a dual to the problem studied in Sections~\ref{sec:def} to~\ref{sec:num}.  We characterize  the asymptotic behavior of $\tau^*(n,\alpha,\varepsilon|\phi_n,P_1,P_2)$ when $\phi_n=\phi_n^{\rm{Gut}}$.

To do so, we recall that the R\'enyi divergence of order $\gamma\in\bbR_+$~\cite{renyi1961measures}  is defined as
\begin{align}
D_{\gamma}(P_1\|P_2):=\frac{1}{\gamma-1}\log \bigg(\sum_{x \in\calX}P_1^{\gamma}(x)P_2^{ 1-\gamma}(x)\bigg).
\end{align}
Note that $\lim_{\gamma\downarrow 1}D_{\gamma}(P_1\|P_2)=D(P_1\|P_2)$, the usual relative entropy.

\begin{proposition}
\label{weakc4bc}
For any $\varepsilon\in(0,1)$, any $\alpha\in\bbR_+$ and any pair of distributions $(P_1,P_2)\in\calP(\calX)^2$, 
\begin{align}
\lim_{n\to\infty}\tau^*(n,\alpha,\varepsilon|\phi_n^{\rm{Gut}},P_1,P_2)=D_{\frac{\alpha}{1+\alpha}}(P_1\|P_2)\label{wcresult}.
\end{align}
\end{proposition}
The proof of Proposition \ref{weakc4bc} is provided in Section \ref{proof:weakc4bc}. Several remarks are in order.

First, the performance of Gutman's test in \eqref{gutmanrule} under this dual setting is dictated by $D_{\frac{\alpha}{1+\alpha}}(P_1\|P_2)$, which is different from $\mathrm{GJS}(P_1,P_2,\alpha)$ in Theorem \ref{bc:second}. Intuitively, this is because  of two reasons. Firstly,  for the type-I error probabilities to be upper bounded by a non-vanishing constant  $\varepsilon\in(0,1)$ for all pairs of distributions, one needs to choose $\lambda = \Theta(\frac{1}{n})$ (implied by  the weak convergence analysis in~\cite{unnikrishnan2016weak}). Consequently, the type-II exponent then satisfies  
\begin{align}
\lim_{\lambda\downarrow 0}
F(P_1,P_2,\alpha,\lambda)&=\min_{Q\in\calP(\calX)}\alpha D(Q\|P_1)+D(Q\|P_2)=D_{\frac{\alpha}{1+\alpha}}(P_1\|P_2).
\end{align}

Second, as $\alpha\to 0$, the exponent $D_{\frac{\alpha}{1+\alpha}}(P_1\|P_2)\to 0$ and thus  the type-II error probability does not decay exponentially fast. However, when $\alpha\to\infty$,  the exponent  $D_{\frac{\alpha}{1+\alpha}}(P_1\|P_2)\to D(P_1\|P_2)$ and thus we can achieve the optimal exponential decay rate of the  type-II error probability as if   $P_1$ and $P_2$ were known (implied by the Chernoff-Stein lemma~\cite{chernoff1952measure}).

Finally, we remark that Proposition \ref{weakc4bc} is not comparable to Theorem \ref{bc:second} since the settings are different. Furthermore, Proposition \ref{weakc4bc} applies only to Gutman's test while Theorem \ref{bc:second} contains an optimization over  {\em all} tests or classifiers.

\section{Classification of Multiple Hypotheses with  the Rejection Option}
\label{sec:result4cmr}
In this section, we generalize our second-order asymptotic result for binary classification in Theorem~\ref{bc:second} to classification of multiple hypotheses with rejection~\cite[Theorem 2]{gutman1989asymptotically}.

\subsection{Problem Formulation}
Given $M$ training sequences $\{X_i^N\}_{i\in[M]}$ generated i.i.d. according to distinct distributions $\{P_i\}_{i\in{M}}\in\calP(\calX)^M$, in classification of multiple hypotheses with rejection, one is asked to determine whether a test sequence $Y^n$ is generated i.i.d. according to a distribution in $\{P_i\}_{i\in[M]}$ or some other distribution. In other words, there are $M+1$ hypotheses:
\begin{itemize}
\item $\rmH_j$ for each $j\in[M]$: the test sequence $Y^n$ and $j^{\mathrm{th}}$ training sequence $X_j^N$ are generated according to the same distribution;
\item $\rmH_\rmr$: the test sequence $Y^n$ is generated according to a distribution different from those in which the training sequences are generated from.
\end{itemize}
In the following, for simplicity, we use $\bX^N$ to denote $(X_1^N,\ldots,X_M^N)$, $\bx^N$ to denote $(x_1^N,\ldots,x_M^N)$ and $\bP$ to denote $(P_1,\ldots,P_M)$. Recall that $N=\alpha n$ for brevity. The main task in classification of multiple hypotheses with rejection is thus to design a test $\psi_n:\calX^{MN}\times\calX^n\to \{\rmH_1,\ldots,\rmH_M,\rmH_\rmr\}$. Note that any such test $\psi_n$ partitions the sample space $\calX^{MN}\times\calX^n$ into $M+1$ disjoint regions: $M$ acceptance regions $\{\calA_j(\psi_n)\}_{j\in[M]}$ where $(\bX^N,Y^n)\in\calA_j(\psi_n)$ favors hypothesis $\rmH_j$ and a rejection region $\calA^\rmc(\psi_n):=\left(\cup_{j\in[M]}\calA_j(\psi_n)\right)^\rmc$ where $(\bX^N,Y^n)\in\calA^\rmc(\psi_n)$ favors hypothesis $\rmH_\rmr$.

Given any test $\psi_n$ and any tuple of distributions $\bP\in\calP(\calX)^M$, we have the following $M$ error probabilities and $M$ rejection probabilities: for each $j\in[M]$,
\begin{align}
\beta_j(\psi_n|\bP)
&:=\bbP_j\big\{\psi_n(\bX^N,Y^n)\notin\{\rmH_j,\rmH_\rmr\}\big\}
\label{def:typejerror},\\
\zeta_j(\psi_n|\bP)
&:=\bbP_j\big\{\psi_n(\bX^N,Y^n)=\rmH_\rmr\big\}
\label{def:typejreject},
\end{align} 
 where similarly to \eqref{def:type1err} and \eqref{def:type2err}, for $j\in[M]$, we define $\bbP_j\{\cdot\}:=\Pr\{\cdot|\rmH_j\}$ where $X_i^N$ is distributed i.i.d.\ according to $P_i$ for all $i\in[M]$.  
We term the probabilities in \eqref{def:typejerror} and \eqref{def:typejreject} as type-$j$ error and rejection probabilities respectively for each $j\in[M]$.

Similarly to Section \ref{sec:results4bc}, we are interested in the following question. For  all tests satisfying (i) for each $j\in[M]$, the type-$j$ error probability decays exponentially fast with the exponent being at least $\lambda\in\bbR_+$ for all tuples of distributions and (ii) for each $j\in[M]$,  the type-$j$ rejection probability is upper bounded by a constant $\varepsilon_j\in(0,1)$ for a particular tuple of distributions, what is the largest achievable exponent $\lambda$? In other words, given $\bm{\varepsilon}=(\varepsilon_1,\ldots,\varepsilon_M)\in(0,1)^M$, we are interested in the following fundamental limit:
\begin{align}
\lambda^*(n,\alpha,\bm{\varepsilon}|\bP)
\nn:=\sup\Big\{\lambda\in\bbR_+: \exists \, \psi_n~\mathrm{s.t.~}\forall j\in[M], \beta_j(\psi_n|\tilde{\bP})&\leq \exp(-n\lambda),\forall~\tilde{\bP}\in\calP(\calX)^M,\\*
\zeta_j(\psi_n|\bP)&\leq \varepsilon_j
\Big\}\label{def:calL}.
\end{align}

\subsection{Main Result}\label{sec:main_M}
For brevity, let $\calM :=\{(r,s)\in[M]^2:r\neq s\}$. Given any  $\bP\in\calP(\calX)^M$, for each $j\in[M]$, let
\begin{align}
\theta_j(\bP,\alpha)
&:=\min_{i\in[M]:i\neq j}\mathrm{GJS}(P_i,P_j,\alpha)\label{def:thetaj}.
\end{align}
Consider any $\bP\in\calP(\calX)^M$ such that the minimizer for $\theta_j(\bP,\alpha)$  in~\eqref{def:thetaj} is unique for each $j\in[M]$ and denote the unique minimizer for $\theta_j(\bP,\alpha)$ as $i^*(j|\bP,\alpha)$. For simplicity, we use $i^*(j)$ to denote $i^*(j|\bP,\alpha)$ when the dependence on $\bP$ is clear.

From Gutman's result in \cite[Thereoms 2 and 3]{gutman1989asymptotically}, we conclude that
\begin{align}
\liminf_{n\to\infty}\lambda^*(n,\alpha,\bm{\varepsilon}|\bP)
&\ge  \min_{j\in[M]}\mathrm{GJS}(P_{i^*(j)},P_j,\alpha)=
 \min_{(i,j)\in\calM}\mathrm{GJS}(P_i,P_j,\alpha).
\end{align}
In this section, we refine the above asymptotic statement, and in particular, derive the second-order approximations to the fundamental limit $\lambda^*(n,\alpha,\bm{\varepsilon}|\bP)$.

Given any tuple of distributions $\bP\in\calP(\calX)^M$ and any vector $\bm{\varepsilon}\in(0,1)^M$, let
\begin{align}
\calJ_1(\bP,\alpha)
&:=\argmin_{j\in[M]}\mathrm{GJS}(P_{i^*(j)},P_j,\alpha)\label{def:calJ1},\\
\calJ_2(\bP,\alpha)
&:=\argmin_{j\in\calJ_1(\bP,\alpha)}\sqrt{\rmV(P_{i^*(j)},P_j,\alpha)}\Phi^{-1}(\varepsilon_j)\label{def:calJ2}.
\end{align}
\begin{theorem}
\label{second:cm}
For any $\alpha\in\bbR_+$, any $\bm{\varepsilon}\in(0,1)^M$ and any tuple of distributions $\bP\in\calP(\calX)^M$ satisfying that the minimizer for $\theta_j(\bP,\alpha)$ is unique for each $j\in[M]$, we have
\begin{align}
\lambda^*(n,\alpha,\bm{\varepsilon}|\bP)
&=\mathrm{GJS}(P_{i^*(j)},P_j,\alpha)+\sqrt{\frac{\rmV(P_{i^*(j)},P_j,\alpha)}{n}}\Phi^{-1}(\varepsilon_j)+O\left(\frac{\log n}{n}\right),  \label{eqn:multiple}
\end{align} 
where \eqref{eqn:multiple} holds for any $j\in\calJ_2(\bP,\alpha)$.
\end{theorem}
 
The proof of Theorem \ref{second:cm} is given in Section \ref{proof:second:cm}. Several remarks are in order.

First, in the achievability proof, we make use of a test proposed by Unnikrishnan~\cite[Theorem~4.1]{unnikrishnan2015asymptotically} and show that it is second-order optimal for classification of multiple hypotheses with  rejection.

Second, we remark that it is not straightforward to obtain the results in Theorem \ref{second:cm} by using the same set of techniques to prove Theorem \ref{bc:second}. The converse proof of Theorem \ref{second:cm} is a generalization of that for Theorem \ref{bc:second}. However, the achievability proof is more involved. As can be gleaned in our proof in Section \ref{proof:second:cm}, the test by Unnikrishnan (see~\eqref{def:unntest}) outputs rejection if the second smallest value of $\{\mathrm{GJS}(\hatT_{X_i^N},\hatT_{Y^n},\alpha)\}_{i\in[M]}$ is smaller than a threshold $\tilde{\lambda}$. The main difficulty lies in identifying the index of the second smallest value in $\{\mathrm{GJS}(\hatT_{X_i^N},\hatT_{Y^n},\alpha)\}_{i\in[M]}$. Note that for each realization of $(\bx^N,y^n)$, such an index can potentially be different. However, we show that for any tuple of distributions $\bP\in\calP(\calX)^M$ satisfying the condition in Theorem \ref{second:cm}, if the training sequences are generated  in an i.i.d.\ fashion according to $\bP$, with probability tending to one, the index of the second smallest value in $\{\mathrm{GJS}(\hatT_{X_i^N},\hatT_{Y^n},\alpha)\}_{i\in[M]}$ under hypothesis $\rmH_j$ is given   by $i^*(j)$. Equipped this important observation, we establish our achievability proof by proceeding similarly to that of Theorem~\ref{bc:second}. 

Finally, we remark that one might also consider tests which provide inhomogeneous performance guarantees under different hypotheses in terms of the error probabilities for all tuples of distributions and, at the same time, constrains the sum of all rejection probabilities to be upper bounded by some $\varepsilon\in(0,1)$. In this direction, the fundamental limit of interest is
\begin{align}
\Lambda(n,\alpha,\varepsilon|\bP)
\nn:=\Big\{\lambda^M\in\bbR_+^M: \exists\ \psi_n~\mathrm{s.t.~}\forall j\in[M], \beta_j(\psi_n|\tilde{\bP})&\leq \exp(-n\lambda_j),~\forall \, \tilde{\bP}\in\calP(\calX)^M,\\*
\sum_{j\in[M]}\zeta_j(\psi_n|\bP)&\leq \varepsilon
\Big\}\label{def:inhomo}.
\end{align}
Characterizing the second-order asymptotics of the set $\Lambda(n,\alpha,\varepsilon|\bP)$ for   $M\geq 3$ is challenging. However, when $M=2$, using similar proof techniques as that for  Theorem \ref{second:cm}, we can characterize the following {\em second-order region}~\cite[Chapter~6]{Tanbook}
\begin{align}
\calL(\alpha,\varepsilon|P_1,P_2)
\nn := \Bigg\{ 
(L_1,L_2)\in\bbR_+ &: \exists\ \{\psi_n\}_{n=1}^\infty \mathrm{~s.t.}~ \forall\, (\tilP_1,\tilP_2)\in\calP(\calX)^2 ,\nn\\*
&\liminf_{n\to\infty}\frac{1}{\sqrt{n}}\Big(  \log \frac{1}{\beta_1(\psi_n|\tilP_1,\tilP_2)} - n\, \mathrm{GJS}(P_1,P_2,\alpha)\Big) \ge L_1,\nn\\*
\nn&\liminf_{n\to\infty}\frac{1}{\sqrt{n}}\Big( \log \frac{1}{\beta_2(\psi_n|\tilP_1,\tilP_2)}  - n\, \mathrm{GJS}(P_2,P_1,\alpha)\Big) \ge L_2,\\*
&\limsup_{n\to\infty}\sum_{j\in[2]}\zeta_j(\psi_n| P_1,P_2)\leq \varepsilon
\Bigg\}\label{def:callseregion}.
\end{align}

Indeed, one can consider the following generalization of Gutman's test~\cite[Theorem 2]{gutman1989asymptotically}
\begin{align}
\psi_n^{\rm{Gut}}(x_1^N,x_2^N,y^n)&:=
\left\{
\begin{array}{ll}
\rmH_1&\mathrm{if~}\mathrm{GJS}(\hatT_{x_2^N},\hatT{y^n},\alpha)-\tilde{\lambda}_2> 0,\\
\rmH_2&\mathrm{if~}\mathrm{GJS}(\hatT_{x_1^N},\hatT_{y^n},\alpha)-\tilde{\lambda}_1> 0,\mathrm{GJS}(\hatT_{x_2^N},\hatT_{y^n},\alpha)-\tilde{\lambda}_2\leq 0\\
\rmH_\rmr&\mathrm{if~}\mathrm{GJS}(\hatT_{x_i^N},\hatT_{y^n},\alpha)-\tilde{\lambda}_i\leq 0,i\in[2]
\end{array}
\right.\label{gutmanbcr},
\end{align}
where $\tilde{\lambda}_1$ and $\tilde{\lambda}_2$ are  thresholds chosen  so that the sum of the type-II error probabilities is upper bounded by $\varepsilon\in(0,1)$. Then, by means of a standard calculation, 
\begin{align}
\calL(\alpha,\varepsilon|P_1,P_2)
&=\bigg\{(L_1,L_2)\in\bbR_+:\Phi\left(\frac{L_1}{\sqrt{\rmV(P_1,P_2,\alpha)}}\right)+\Phi\left(\frac{L_2}{\sqrt{\rmV(P_2,P_1,\alpha)}}\right)\leq \varepsilon\bigg\}.
\end{align}
This result clearly elucidates a trade-off between   $L_1$ and $L_2$ or, equivalently, the two  rejection probabilities  $\zeta_1(\psi_n| P_1,P_2)$ and $\zeta_2(\psi_n| P_1,P_2)$.

\subsection{Analysis in A Dual Setting}
\label{sec:dual:cm}

Similar to the analysis of the dual setting in Section~\ref{sec:weakconvergence}, for classification of multiple hypotheses with the rejection option, one might be interested in studying tests whose type-$j$ error probability for each $j\in[M]$ are upper bounded by a constant for all tuples of distributions $\tilde{\bP}\in\calP(\calX)^M$ and whose type-$j$ rejection probability for each $j\in[M]$ decays exponentially fast for a particular  $\bP\in\calP(\calX)^M$. To be specific, given any decision rule $\Psi_n$ and any $\varepsilon\in(0,1)$, we study the following non-asymptotic fundamental limit:
\begin{align}
\tau^*(n,\alpha,\varepsilon|\Psi_n,\bP)
\nn:=\sup\big\{\tau\in\bbR_+: \exists\, \psi_n~\mathrm{s.t.~}\forall j\in[M], \beta_j(\psi_n|\tilde{\bP})&\leq \varepsilon,\forall~\tilde{\bP}\in\calP(\calX)^M,\\*
\zeta_j(\psi_n|\bP)&\leq \exp(-n\tau)
\big\}\label{def:caltau2}.
\end{align}
To analyze the fundamental limit in \eqref{def:caltau2}, given   training and test sequences $(\bx^M,y^n)$, we consider Gutman's test~\cite[Theorem 2]{gutman1989asymptotically} which  is given by the following rule
\begin{align}
\Psi_n^{\rm{Gut}}
(\bx^M,y^n)
&:=
\left\{
\begin{array}{ll}
\rmH_1&\mathrm{if~}\max_{i\in[M]:i\neq 1}\mathrm{GJS}(\hatT_{x_i^N},\hatT_{y^n},\alpha)>\lambda,\\
\rmH_j&\mathrm{if~}\max_{i\in[M]:i\neq j}\mathrm{GJS}(\hatT_{x_i^N},\hatT_{y^n},\alpha)>\lambda,\mathrm{GJS}(\hatT_{x_j^N},\hatT_{y^n},\alpha)\leq \lambda,\\
\rmH_\rmr& \mathrm{otherwise}
\end{array}
\right.\label{gut:mwithreject}
\end{align} 
for $j\in[2:M]$. 
The reason why, unlike in Section~\ref{sec:main_M},  we do not analyze Unnikrishnan's test~\cite{unnikrishnan2015asymptotically}  (see~\eqref{def:unntest}) is because it is designed so that the $j$-th error probability  $ \beta_j(\psi_n|\tilde{\bP}) $  decays exponentially fast for every tuple of distributions $\tilde{\bP}$. Since \eqref{def:caltau2} stipulates that  $\beta_j(\psi_n|\tilde{\bP}) $ is non-vanishing, clearly Unnikrishnan's test is not suited to this dual regime. 

To present our result, we need the following definition.  Given any triple of distributions $(P_1,P_2,P_3)\in\calP(\calX)^3$ and any $\gamma\in\bbR_+$, define a generalized divergence measure between three distributions as
\begin{align}
D_{\gamma}(P_1,P_2,P_3)
&:=\frac{1}{\gamma-1}\log\bigg(\sum_{x}P_1(x)^{1-\gamma}P_2(x)^{\frac{\gamma}{2}}P_3(x)^{\frac{\gamma}{2}}\bigg)\label{def:gdivergence}.
\end{align}

\begin{proposition}
\label{dual:unn}
For any $\alpha\in\bbR_+$, any $\varepsilon\in(0,1)$ and any tuple of distributions $\bP\in\calP(\calX)^M$, we have
\begin{align}
\lim_{n\to\infty}\tau^*(n,\alpha,\varepsilon|\Psi_n^{\rm{Gut}},\bP)&=\min_{j\in[M]}\min_{(i,k)\in\calM}D_{\frac{2\alpha}{1+2\alpha}}(P_j,P_i,P_k),
\end{align}
where $\calM=\{(r,s)\in[M]^2:r\neq s\}$.
\end{proposition}
The proof of Proposition \ref{dual:unn} is provided in Section \ref{proof:dual:unn}. Several remarks are in order.

 First, the exponent of Gutman's test in~\eqref{gut:mwithreject} in  the dual setting is considerably different from that in Theorem~\ref{second:cm}. Intuitively, this is because for this setting, in order to ensure that the error probability under each hypothesis is upper bounded by $\varepsilon$ for all $\tilde{\bP}$, we need to choose $\lambda=\Theta(\frac{1}{n})$ in \eqref{gut:mwithreject}. In contrast, $\lambda$ is chosen to $\Theta(1)$ in the proof of Theorem \ref{second:cm}.

 Second, as $\alpha\to 0$, the exponent $D_{\frac{2\alpha}{1+2\alpha}}(P_j,P_i,P_k)\to 0$ for each $(j,i,k)\in[M]\times\calM$. Thus if the ratio of the lengths of the training to test sequences is vanishingly small, the rejection probabilities cannot decay exponentially fast. This conforms to our intuition as there are too few  training samples to  train effectively.  

 Finally, when $\alpha\to\infty$,  one can verify that $D_{\frac{2\alpha}{1+2\alpha}}(P_j,P_i,P_k)\to \infty$ and thus the rejection probabilities decay super exponentially fast if the length of the training sequences $N$ is scaling faster than the length of the test sequence $n$, i.e., $N=\omega(n)$.  In contrast, in Proposition~\ref{weakc4bc}, when $\alpha\to \infty$, the exponent of type-II error probability for any $(P_1,P_2)$ converges to the Chernoff-Stein exponent $D(P_1\|P_2)$~\cite{chernoff1952measure}, which is finite. Why is there a dichotomy when in both settings,  $N$ is much larger than  $n$ and so one can estimate the underlying distributions with arbitrarily high accuracy? The dichotomy between these two results is due to a subtle difference  in two settings, which we delineate here. In Proposition \ref{weakc4bc}, a test sequence is generated  according to $P_1$ or $P_2$ and one is asked to make a decision {\em without  the rejection option}.  If the true pair of distributions is known, the setting basically reduces to {\em binary hypothesis testing}~\cite{chernoff1952measure} and so $D(P_1\| P_2)$ is the type-II exponent. However, in Proposition~\ref{dual:unn}, a test sequence is generated  according to one of the $M$ unknown distributions in $\bP$ and one is also {\em allowed the rejection option}. When the true $\bP$ is known (i.e., the case $\alpha\to\infty$ which allows one to estimate $\bP$ accurately), the setting in Proposition~\ref{dual:unn} essentially reduces to  {\em $M$-ary hypothesis testing} in which rejection is no longer permitted, which implies that the exponent of the probability of the rejection event  $\tau^*(n,\alpha,\varepsilon|\Psi_n^{\mathrm{Gut}},\bP)$ tends to infinity. 
 


\section{Proof of the Main Results}
\label{sec:proofs}
\subsection{Proof Theorem \ref{bc:second}}
\label{proof:bcsecond}
In this section, we present the proof of second-order asymptotics for the binary classification problem. The main techniques used are the method of types, Taylor approximations of the generalized Jensen-Shannon divergence and a careful application of the central limit theorem.

\subsubsection{Achievability Proof}
\label{sec:ach4bc}
In the achievability proof, we use Gutman's test \eqref{gutmanrule} with the threshold $\lambda$ replaced by
\begin{align}
\tilde{\lambda}&:=\lambda-\frac{|\calX|\log \big((1+\alpha)n+1\big)}{n}\label{def:tildelambda}.
\end{align}
Given any  $(P_1,P_2)\in\calP(\calX)^2$, the type-I and type-II error probabilities for $\phi_n^{\rm{Gut}}$ are given by
\begin{align}
\beta_1(\phi_n^{\rm{Gut}}|P_1,P_2)
&=\bbP_1\Big\{\mathrm{GJS}(\hatT_{X_1^N},\hatT_{Y^n},\alpha)>\tilde{\lambda}\Big\},\\*
\beta_2(\phi_n^{\rm{Gut}}|P_1,P_2)
&=\bbP_2\Big\{\mathrm{GJS}(\hatT_{X_1^N},\hatT_{Y^n},\alpha)\leq \tilde{\lambda}\Big\}.
\end{align}

We first analyze $\beta_2(\phi_n^{\rm{Gut}}|P_1,P_2)$. Given any $P\in\calP(\calX)$, define the following typical set:
\begin{align}
\calB_n(P)
&:=\bigg\{x^n\in\calX^n:\max_{x\in\calX}|\hatT_{x^n}(x)-P(x)|\leq \sqrt{\frac{\log n}{n}}\bigg\}\label{def:typical}.
\end{align}
By Chebyshev's inequality (see also \cite[Lemma 22]{tan2014state}), we can show that
\begin{align}
\bbP_2\Big\{X_1^N\notin\calB_N(P_1)~\mathrm{or}~Y^n\notin\calB_n(P_2)\Big\}
&\leq \frac{2|\calX|}{N^2}+\frac{2|\calX|}{n^2}=\frac{2(1+\alpha^2)|\calX|}{2\alpha^2 n^2}=:\tau_n\label{pofatypical}.
\end{align}

Recall the definitions of information densities in \eqref{def:i}. It is easy to verify that
\begin{align}
\mathrm{GJS}(P_1,P_2,\alpha)
&=\alpha\mathbb{E}_{P_1}\left[\imath_1(X|P_1,P_2,\alpha)\right]+\mathbb{E}_{P_2}\left[\imath_2(X|P_1,P_2,\alpha)\right]\label{averageimath=GJS}.
\end{align}
Furthermore, for any pair of distributions $(P_1,P_2)\in\calP(\calX)^2$ and any $\alpha\in\bbR_+$, the derivatives of the generalized Jensen-Shannon divergence $\mathrm{GJS}(P_1,P_2,\alpha)$   are as follows:
\begin{alignat}{2}
\frac{\partial \mathrm{GJS}(P_1,P_2,\alpha)}{\partial P_1(x)}
&=\alpha\imath_1(x|P_1,P_2,\alpha),   \qquad &\forall\, x\in\supp(P_1)\label{firstd1},\\
\frac{\partial \mathrm{GJS}(P_1,P_2,\alpha)}{\partial P_2(x)}
&=\imath_2(x|P_1,P_2,\alpha),  \qquad &   \forall\, x\in\supp(P_2)\label{firstd2},\\
\frac{\partial^2 \mathrm{GJS}(P_1,P_2,\alpha)}{\partial (P_1(x))^2}
&=\frac{\alpha P_2(x)}{P_1(x)(\alpha P_1(x)+P_2(x))},\qquad &\forall\, x\in\supp(P_1),\\
\frac{\partial^2 \mathrm{GJS}(P_1,P_2,\alpha)}{\partial (P_2(x))^2}
&=\frac{\alpha P_1(x)}{P_2(x)(\alpha P_1(x)+P_2(x))},\qquad &\forall\, x\in\supp(P_2),\\
\frac{\partial^2 \mathrm{GJS}(P_1,P_2,\alpha)}{\partial P_1(x)P_2(x)}
&=-\frac{\alpha}{\alpha P_1(x)+P_2(x)},\qquad &\forall\, x\in\supp(P_1)\cap\supp(P_2) \label{seconddlast}.
\end{alignat}

Using the results in \eqref{firstd1}--\eqref{seconddlast} and applying a Taylor expansion to $\mathrm{GJS}(\hatT_{x_1^N},\hatT_{y^n},\alpha)$ around $(P_1,P_2)$ for any $x_1^N\in\calB_N(P_1)$ and $y^n\in\calB_n(P_2)$, we obtain 
\begin{align}
\nn&\mathrm{GJS}(\hatT_{x_1^N},\hatT_{y^n},\alpha)\\*
\nn&=\mathrm{GJS}(P_1,P_2,\alpha)+\sum_{x\in\calX}(\hatT_{x_1^N}(x)-P_1(x))\alpha\imath_1(x|P_1,P_2,\alpha)+\sum_{x\in\calX}(\hatT_{y^n}(x)-P_2(x))\imath_2(x|P_1,P_2,\alpha)\\*
&\qquad+O(\|\hatT_{x_1^N}-P_1\|^2+O(\|\hatT_{y^n}-P_2\|^2))\\*
&=\frac{1}{n}\sum_{i\in[N]}\imath_1(x_{1,i}|P_1,P_2,\alpha)
+\frac{1}{n}\sum_{i\in[n]}\imath_2(y_i|P_1,P_2,\alpha)+O\left(\frac{\log n}{n}\right)\label{taylor},
\end{align}
where \eqref{taylor} follows because $N= \lceil n\alpha \rceil$ and the fact that the types in $\calB_N(P_1)$ and $\calB_n(P_2)$ are $O(\sqrt{\frac{\log n}{n}})$-close to the generating (underlying) distributions $P_1$ and $P_2$.

Recall the definition  of $\rmV(P_1,P_2,\alpha)$ in \eqref{def:v}. Let the linear combination of the third absolute moments of the information densities in \eqref{def:i} be defined as
\begin{align}
\rmT(P_1,P_2,\alpha)
&:=\alpha\mathbb{E}_{P_1}\left[\big|\imath_1(X|P_1,P_2,\alpha)-\mathbb{E}_{P_i}[\imath_i(X|P_1,P_2,\alpha)]\big|^3\right] \nn\\*
&\qquad+\mathbb{E}_{P_2}\left[\big|\imath_2(X|P_1,P_2,\alpha)-\mathbb{E}_{P_2}[\imath_2(X|P_1,P_2,\alpha)]\big|^3\right]\label{def:rmT}.
\end{align}

We can upper bound the type-II error probability as follows:
\begin{align}
\nn&\beta_2(\phi_n^{\rm{Gut}}|P_1,P_2)=\bbP_2\Big\{\mathrm{GJS}(\hatT_{X_1^N},\hatT_{Y^n},\alpha)\leq \tilde{\lambda}\Big\}\\*
\nn&\le\bbP_2\Big\{\mathrm{GJS}(\hatT_{X_1^N},\hatT_{Y^n},\alpha)\leq\tilde{\lambda},X_1^N\in\calB_N(P_1),Y^n\in\calB_n(P_2)\Big\}\\*
&\qquad+\bbP_2\Big\{X_1^N\notin\calB_N(P_1)\mathrm{~or~}Y^n\notin\calB_n(P_2)\Big\}\\
&=\bbP_2\bigg\{\frac{1}{n}\sum_{i\in[N]}\imath_1(X_{1,i}|P_1,P_2,\alpha)
+\frac{1}{n}\sum_{i\in[n]}\imath_2(Y_i|P_1,P_2,\alpha)+O\left(\frac{\log n}{n}\right)\leq \tilde{\lambda}\bigg\}+\tau_n\label{usetaylor&atypical}\\
\nn&=\bbP_2\bigg\{\frac{1}{n+N}\sum_{i\in[N]}\big(\imath_1(X_{1,i}|P_1,P_2,\alpha)-\mathbb{E}_{P_1}[\imath_1(x|P_1,P_2,\alpha)]\big) \nn\\*
&\quad +\frac{1}{n\!+\! N}\sum_{i\in[n]}\big(\imath_2(Y_i|P_1,P_2,\alpha)\!-\!\mathbb{E}_{P_2}[\imath_2(x|P_1,P_2,\alpha)]\big)\!\leq \!\frac{\lambda-\mathrm{GJS}(P_1,P_2,\alpha)+O (\frac{ \log n}{n} ) }{1+\alpha}\bigg\}\!+\!\tau_n\label{usel&ei}\\
&\leq \Phi\left(\left(\lambda-\mathrm{GJS}(P_1,P_2,\alpha)+O\left(\frac{\log n}{n}\right)\right)\sqrt{\frac{n}{\rmV(P_1,P_2,\alpha)}}\right)+\frac{6\rmT(P_1,P_2,\alpha)}{\sqrt{n(\rmV(P_1,P_2,\alpha))^3}}+\tau_n\label{useberry},
\end{align}
where \eqref{usetaylor&atypical} follows from the bound in~\eqref{pofatypical} and the Taylor expansion in~\eqref{taylor}; \eqref{usel&ei} follows from the expression for $\mathrm{GJS}(P_1,P_2,\alpha)$  in \eqref{averageimath=GJS}, the fact that $N=n\alpha$ and the definition of $\tilde{\lambda}$ in \eqref{def:tildelambda}; and \eqref{useberry} follows from the Berry-Esseen theorem~\cite{berry1941accuracy,esseen1942}.

Similarly to \eqref{def:type1err} and \eqref{def:type2err}, for $j\in[2]$, we define $\tilde{\bbP}_j\{\cdot\}:=\Pr\{\cdot|\rmH_j\}$ where $(X_1^N,X_2^N)$ are generated from the pair of distributions $(\tilP_1,\tilP_2)$. For all  $(\tilP_1,\tilP_2)\in\calP(\calX)^2$, the type-I error probability can be upper bounded as follows:
\begin{align}
\beta_1(\phi_n^{\rm{Gut}}|\tilP_1,\tilP_2)
&=\tilde{\bbP}_1\Big\{\mathrm{GJS}(\hatT_{X_1^N},\hatT_{Y^n},\alpha)>\tilde{\lambda}\Big\}\\*
&=\sum_{x_1^N,y^n: \mathrm{GJS}(\hatT_{x_1^N},\hatT_{y^n},\alpha)>\tilde{\lambda}}\tilP_1^N(x_1^N)\tilP_1^n(y^n)\\
&=\sum_{ (Q_1,Q_2): \mathrm{GJS}(Q_1,Q_2,\alpha)> \tilde{\lambda}} \tilP_1^N(\calT^N_{Q_1})\tilP_1^n(\calT^n_{Q_2})\\
&\leq \sum_{(Q_1,Q_2) :\mathrm{GJS}(Q_1,Q_2,\alpha)\geq \tilde{\lambda}}\exp\big\{-ND(Q_1\|\tilP_1)-nD(Q_2\|\tilP_1)\big\}\\
&\leq \sum_{ (Q_1,Q_2): \mathrm{GJS}(Q_1,Q_2,\alpha)\geq \tilde{\lambda}}\exp(-n\tilde{\lambda})\exp\bigg\{-n(1+\alpha)D\left(\frac{\alpha Q_1+Q_2}{1+\alpha}\Big\|\tilP_1\right)\bigg\}\label{explain1}\\
&\leq \exp(-n\tilde{\lambda})\sum_{Q\in\calP_{n+N}(\calX)}\exp\big\{-(n+N)D(Q\|\tilP_1)\big\}\\
&\leq \exp(-n\tilde{\lambda})\sum_{Q\in\calP_{n+N}(\calX)}(n+N+1)^{|\calX|}\tilP_1^{n+N}(\calT^{n+N}_{Q})\\
&\leq \exp\Big\{-n\tilde{\lambda}+|\calX|\log \big((1+\alpha)n+1\big)\Big\}\\*
&=\exp(-n\lambda),\label{upptype2}
\end{align}
where \eqref{upptype2} follows from the definition of $\tilde{\lambda}$ in \eqref{def:tildelambda} and \eqref{explain1} follows since
\begin{align}
\nn&ND(Q_1\|\tilP_1)+nD(Q_2\|\tilP_1)\\*
&=n\alpha\mathbb{E}_{Q_1}\left[\log\frac{Q_1(X)}{\tilP_1(X)}\right]+n\mathbb{E}_{Q_2}\left[\log\frac{Q_2(X)}{\tilP_1(X)}\right]\\
&=n\alpha\mathbb{E}_{Q_1}\left[\log\frac{(1+\alpha)Q_1(X)}{\alpha Q_1(X)+Q_2(X)}\right] +n\mathbb{E}_{Q_2}\left[\log\frac{(1+\alpha)Q_2(X)}{\alpha Q_1(X)+Q_2(X)}\right]\nn\\*
&\qquad+n(1+\alpha)D\left(\frac{\alpha Q_1+Q_2}{1+\alpha}\Big\|\tilP_1\right)\\
&=n\mathrm{GJS}(Q_1,Q_2,\alpha)+n(1+\alpha)D\left(\frac{\alpha Q_1+Q_2}{1+\alpha}\Big\|\tilP_1\right)\\
&\geq n\lambda+n(1+\alpha)D\left(\frac{\alpha Q_1+Q_2}{1+\alpha}\Big\|\tilP_1\right).
\end{align}
For brevity, let
\begin{align}
\rho_n&:=\frac{6\rmT(P_1,P_2,\alpha)}{\sqrt{n(\rmV(P_1,P_2,\alpha))^3}}+\tau_n\label{def:rhon}.
\end{align}

Combining the results in \eqref{useberry} and \eqref{upptype2}, if we choose $\lambda\in\bbR_+$ s.t.,
\begin{align}
\lambda
&=\mathrm{GJS}(P_1,P_2,\alpha)+\sqrt{\frac{\rmV(P_1,P_2,\alpha)}{n}}\Phi^{-1}\left(\varepsilon-\rho_n\right)+O\left(\frac{\log n}{n}\right),
\end{align}
Gutman's test with threshold $\tilde{\lambda}$ in \eqref{def:tildelambda} satisfies that i) $\beta_1(\phi_n^{\rm{Gut}}|\tilP_1,\tilP_2)\leq \exp(-n\lambda)$ for all $(\tilP_1,\tilP_2)\in\calP(\calX)^2$, and (ii) $\beta_2(\phi_n^{\rm{Gut}}|P_1,P_2)\leq \varepsilon$.
Therefore, we conclude that
\begin{align}
\lambda^*(n,\alpha,\varepsilon|P_1,P_2)
&\geq \mathrm{GJS}(P_1,P_2,\alpha)+\sqrt{\frac{\rmV(P_1,P_2,\alpha)}{n}}\Phi^{-1}(\varepsilon-\rho_n)+O\left(\frac{\log n}{n}\right)\\
&=\mathrm{GJS}(P_1,P_2,\alpha)+\sqrt{\frac{\rmV(P_1,P_2,\alpha)}{n}}\Phi^{-1}(\varepsilon)+O\left(\frac{\log n}{n}\right)\label{taylorphi},
\end{align}
where \eqref{taylorphi} follows from a Taylor approximation of $\Phi^{-1}(\cdot)$ (cf. \cite[Corollary 51]{polyanskiy2010finite}).

\subsubsection{Converse Proof}
\label{sec:converse4bc}
The following lemma relates the error probabilities of any   test to a type-based test (i.e., a test which is a function of only the marginal types $(\hatT_{X_1^N}, \hatT_{X_2^N}, \hatT_{Y^n})$).
\begin{lemma}
\label{anytotype}
For any arbitrary test $\phi_n$, given any $\kappa\in[0,1]$ and any pair of distributions $(P_1,P_2)\in\calP(\calX)^2$, we can construct a type-based test $\phi_n^\rmT$ such that
\begin{align}
\beta_1(\phi_n|P_1,P_2)&\geq \kappa \beta_1(\phi_n^\rmT|P_1,P_2),\\*
\beta_2(\phi_n|P_1,P_2)&\geq (1-\kappa)\beta_2(\phi_n^\rmT|P_1,P_2).
\end{align}
\end{lemma}
The proof of Lemma \ref{anytotype} is inspired by \cite[Lemma 2]{gutman1989asymptotically} and provided in Appendix \ref{proof:anytotype}.

The following lemma shows that for any type-based test $\phi_n^\rmT$, if we constrain the type-I error probability to decay exponentially fast for all pairs of distributions, then the type-II error probability for any particular pair of distributions can be lower bounded by a certain cdf of the generalized Jensen-Shannon divergence evaluated at the marginal types of the training and test sequences. The lemma can be used to assert that Gutman's test in~\eqref{gutmanrule} is ``almost'' optimal when restricted to the class of all type-based tests. For brevity, given $(\alpha,t)\in\bbR_+^2$, let
\begin{align}
\eta_n(\alpha)&:=\frac{|\calX|\log (n+1)}{n} + \frac{2|\calX|\log (1+\alpha n)}{\alpha n}\label{def:etan}.
\end{align}

\begin{lemma}
\label{typeconverse}
For any $\lambda\in\bbR_+$ and any type-based test $\phi_n^\rmT$ satisfying that for all pairs of distributions $(\tilP_1,\tilP_2)\in\calP(\calX)^2$,
\begin{align}
\beta_1(\phi_n^\rmT|\tilP_1,\tilP_2)\leq \exp(-n\lambda),\label{typeconstraint}
\end{align}
we have that for any pair of distributions $(P_1,P_2)\in\calP(\calX)^2$,
\begin{align}
\beta_2(\phi_n^\rmT|P_1,P_2)&\geq \bbP_2\Big\{\mathrm{GJS}(\hatT_{X_1^N},\hatT_{Y^n},\alpha)+\eta_n(\alpha)<\lambda\Big\}.
\end{align}
\end{lemma}
The proof of Lemma \ref{typeconverse} is inspired by \cite[Theorem 1]{gutman1989asymptotically} and provided in Appendix \ref{proof:typeconverse}.

Combining the results in Lemmas \ref{anytotype} and \ref{typeconverse} and  letting $\kappa={1}/{n}$, we obtain the following corollary.
\begin{corollary}
\label{converse}
Given any $\lambda\in\bbR_+$, for any test $\phi_n$ satisfying the condition that for all pairs of distributions $(\tilP_1,\tilP_2)\in\calP(\calX)^2$
\begin{align}
\beta_1(\phi_n|\tilP_1,\tilP_2)\leq \exp(-n\lambda)\label{testconstraint},
\end{align}
we have that any pair of distributions $(P_1,P_2)\in\calP(\calX)^2$, 
\begin{align}
\beta_2(\phi_n|P_1,P_2)\geq \left(1-\frac{1}{n}\right)\bbP_2\Big\{\mathrm{GJS}(\hatT_{X_1^N},\hatT_{Y^n},\alpha)+\eta_n(\alpha)+\frac{\log n}{n}<\lambda\Big\}.
\end{align}
\end{corollary}

Using Corollary \ref{converse}, the converse part of our second-order asymptotics can be proved similarly to the achievability part by using the result in \eqref{pofatypical}, the Taylor expansions in~\eqref{taylor}, the definition of $\rho_n$ in~\eqref{def:rhon} and applying the Berry-Esseen theorem similarly to \eqref{useberry}. Invoking Corollary~\ref{converse}, we obtain that for any test $\phi_n$ satisfying \eqref{testconstraint} and any pair of distributions $(P_1,P_2)\in\calP(\calX)^2$,
\begin{align}
\nn&\beta_2(\phi_n|P_1,P_2) \geq \Big(1-\frac{1}{n}\Big)\bbP_2\Big\{\mathrm{GJS}(\hatT_{X_1^N},\hatT_{Y^n},\alpha)+O\left(\frac{\log n}{n}\right)<\lambda\Big\}\\
&\quad\geq \Big(1-\frac{1}{n}\Big)\bbP_2\Big\{\mathrm{GJS}(\hatT_{X_1^N},\hatT_{Y^n},\alpha)+O\left(\frac{\log n}{n}\right)<\lambda,X_1^N\in\calB_N(P_1),Y^n\in\calB_n(P_2)\Big\}\\
\nn&\quad\geq\Big(1-\frac{1}{n}\Big)\bbP_2\bigg\{\frac{1}{n}\sum_{i\in[N]}\imath_1(X_{1,i}|P_1,P_2,\alpha)+\frac{1}{n}\sum_{i\in[n]}\imath_2(Y_i|P_1,P_2,\alpha)+O\left(\frac{\log n}{n}\right)<\lambda\bigg\}\\*
&\qquad-\Big(1-\frac{1}{n}\Big)\bbP_2\Big\{X_1^N\notin\calB_N(P_1)~\mathrm{or}~Y^n\notin\calB_n(P_2)\Big\}\\
&\quad\geq \Big(1-\frac{1}{n}\Big)\bigg\{\Phi\left(\left(\lambda-\mathrm{GJS}(P_1,P_2,\alpha)+O\left(\frac{\log n}{n}\right)\right)\sqrt{\frac{n}{\rmV(P_1,P_2,\alpha)}}\right)-\rho_n\bigg\}\label{converse:step1}.
\end{align}

Using \eqref{converse:step1} and the definition of $\lambda^*(\cdot|\cdot)$ in \eqref{def:l1^*}, we conclude that for any  $(P_1,P_2)\in\calP(\calX)^2$,
\begin{align}
\lambda^*(n,\alpha,\varepsilon|P_1,P_2)
&\le\mathrm{GJS}(P_1,P_2,\alpha)+\sqrt{\frac{\rmV(P_1,P_2,\alpha)}{n}}\Phi^{-1}(\varepsilon)+O\left(\frac{\log n}{n}\right),
\end{align}
where a Taylor approximation of $\Phi^{-1}(\cdot)$ has been applied.

\subsection{Proof of Proposition \ref{weakc4bc}}
\label{proof:weakc4bc}
\subsubsection{Preliminaries}
In this subsection, we recall a weak convergence result of Unnikrishnan and Huang~\cite{unnikrishnan2016weak}  and present a key lemma for the   analysis of Gutman's decision rule in \eqref{gutmanrule}.

Under   $\rmH_1$, for all pairs of distributions $(\tilP_1,\tilP_2)\in\calP(\calX)^2$, the weak convergence result in Unnikrishnan and Huang~\cite[Lemma 5]{unnikrishnan2016weak} shows that 
\begin{align}
2n \mathrm{GJS}(\hatT_{X_1^N},\hatT_{Y^n},\alpha)\stackrel{\rmd}{\longrightarrow}\chi^2_{|\calX|-1}\label{weakconjh}.
\end{align}

The following properties of  $F(P_1,P_2,\alpha,\lambda)$, defined in~\eqref{def:FP1P2l}, play an important role in our analyses.
\begin{lemma}
\label{propF}
The type-II exponent function $F(P_1,P_2,\alpha,\lambda)$ satisfies that $F(P_1,P_2,\alpha,0)=D_{\frac{\alpha}{1+\alpha}}(P_1\|P_2)$ and the distribution achieving $F(P_1,P_2,\alpha,0)$ is $Q^*=P^{(\frac{\alpha}{1+\alpha})}$, where $P^{(\gamma)}$ is the {\em tilted distribution}
\begin{equation}
P^{(\gamma)}(x):=\frac{P_1^{\gamma}(x)P_2(x)^{1-\gamma}}{\sum_{a\in\calX} P_1^{\gamma}(a)P_2^{1-\gamma}(a)}\label{def:Pgamma}.
\end{equation}
\end{lemma}
The proof of Lemma \ref{propF} follows directly from applying the KKT conditions~\cite{boyd2004convex}  to   $F(P_1,P_2,\alpha,0)$, defined in \eqref{def:FP1P2l}, and so it is omitted.

\subsubsection{Achievability Proof}
Recall Gutman's test $\phi_n^{\rm{Gut}}$ in \eqref{gutmanrule}. Also recall that $\rmG_k^{-1}(\cdot)$ is the inverse of the complementary cdf of a chi-square random variable with $k$ degrees of freedom. If we choose
\begin{align}
\lambda=\frac{1}{2n}\rmG_{|\calX|-1}^{-1}(\varepsilon)\label{lambddainwca},
\end{align}
then using \eqref{weakconjh} and letting $Z\sim\chi^2_{|\calX|-1}$, we have that for all  $(\tilP_1,\tilP_2)\in\calP(\calX)^2$,
\begin{align}
\limsup_{n\to\infty}\beta_1(\phi_n^{\rm{Gut}}|\tilP_1,\tilP_2)
&=\limsup_{n\to\infty}\tilde{\bbP}_1\Big\{\mathrm{GJS}(\hatT_{X_1^N},\hatT_{Y^n},\alpha)> \lambda\Big\}=\Pr\left\{Z>\rmG_{|\calX|-1}^{-1}(\varepsilon)\right\}=\varepsilon\label{needtouse}.
\end{align}
Furthermore, following similar steps as in \cite{gutman1989asymptotically}, for any $(P_1,P_2)\in\calP(\calX)^2$, we can upper bound the type-II error probability as follows
\begin{align}
\beta_2(\phi_n^{\rm{Gut}}|P_1,P_2)
&\leq (n+1)^{|\calX|}(N+1)^{|\calX|}\exp\{-nF(P_1,P_2,\alpha,\lambda)\}\label{methodoftypeseasy}.
\end{align}
Using Lemma \ref{propF} and the fact that $F(P_1,P_2,\alpha,\lambda)$ is continuous in $\lambda$ \cite[Lemma~12]{Tan11_IT}, we obtain that 
\begin{align}
\liminf_{n\to\infty}-\frac{1}{n}\log \beta_2(\phi_n^{\rm{Gut}}|P_1,P_2)
&\geq \liminf_{n\to\infty} F\left(P_1,P_2,\alpha,\frac{\rmG_{|\calX|-1}^{-1}(\varepsilon)}{2n}\right)\\
&=D_{\frac{\alpha}{1+\alpha}}(P_1\|P_2).
\end{align}

\subsubsection{Converse Proof for Gutman's Test}
From the result in \eqref{needtouse}, we conclude that in order for Gutman's test to satisfy that
\begin{align}
\limsup_{n\to\infty}\beta_1(\phi_n^{\rm{Gut}}|\tilP_1,\tilP_2)\leq \varepsilon,~\forall~(\tilP_1,\tilP_2)\in\calP(\calX)^2,
\end{align}
the threshold $\lambda$ in Gutman's test in \eqref{gutmanrule} should satisfy that
\begin{align}
\lambda\geq \frac{1}{2n}\rmG_{|\calX|-1}^{-1}(\varepsilon)\label{needtouse2}.
\end{align}

For simplicity, similar to \eqref{def:FP1P2l}, let
\begin{align}
F_n(P_1,P_2,\alpha,\lambda)
&:=\min_{\substack{(Q_1,Q_2)\in\calP_N(\calX)\times\calP_n(\calX):\\\mathrm{GJS}(Q_1,Q_2,\alpha)\leq \lambda}} \alpha D(Q_1\|P_1)+D(Q_2\|P_2)\label{def:Fn}.
\end{align}

Using the decision rule in \eqref{gutmanrule}, for any  $(P_1,P_2)\in\calP(\calX)^2$, we can lower bound the type-II error probability as follows:
\begin{align}
&\beta_2(\phi_n^{\rm{Gut}}|P_1,P_2)=\bbP_2\Big\{\phi_n^{\rm{Gut}}(Y^n,X_1^N,X_2^N)=\rmH_1\Big\}\\
&\quad=\sum_{(Q_1,Q_2) : \mathrm{GJS}(Q_1,Q_2,\alpha)\leq \lambda} P_2^n(\calT^n_{Q_2})P_1^N(\calT^N_{Q_1})\\
&\quad\geq \sum_{(Q_1,Q_2):\mathrm{GJS}(Q_1,Q_2,\alpha)\leq \lambda} (n+1)^{-|\calX|}(N+1)^{-|\calX|}\exp\big(-ND(Q_1\|P_1)-nD(Q_2\|P_2)\big)\\
&\quad\geq (n+1)^{-|\calX|}(N+1)^{-|\calX|}\exp(-nF_n(P_1,P_2,\alpha,\lambda))\\
&\quad\geq \exp\big(-nF_n(P_1,P_2,\alpha,0)-|\calX|\log (n+1)-|\calX|\log(n\alpha+1)\big)\label{decreasinl}.
\end{align}
where \eqref{decreasinl} follows since $\lambda\geq 0$ (see \eqref{needtouse2} and $F_n(P_1,P_2,\alpha,\lambda)$ is non-increasing in  $\lambda$. The proof of the converse is completed by invoking the following lemma which relates $F_n(P_1,P_2,\alpha,0)$ to $F(P_1,P_2,\alpha,0)=D_{\frac{\alpha}{1+\alpha}}(P_1\|P_2)$. For brevity, let $n':=\min\{n,N\}=\min\{n, \lceil n\alpha\rceil\}$.
\begin{lemma}
\label{fn<=f+}
For any $(P_1,P_2)\in\calP(\calX)^2$ and any $\alpha\in\bbR_+$, we have
\begin{align}
F_n(P_1,P_2,\alpha,0)
&\leq D_{\frac{\alpha}{1+\alpha}}(P_1\|P_2)+\frac{(1+\alpha)|\calX|}{n'}\log n'-\frac{\sum_x \log (P_1^\alpha(x)P_2(x))}{n'}.
\end{align}
\end{lemma}
The proof of Lemma \ref{fn<=f+} is provided in Appendix \ref{proof:fn<=f+}.

\subsection{Proof of Theorem \ref{second:cm}}
\label{proof:second:cm}
We present the proof for the second-order asymptotics for classification of multiple hypotheses with rejection.
\subsubsection{Achievability Proof}
We use a test proposed by Unnikrishnan in~\cite[Theorem~4.1]{unnikrishnan2015asymptotically}. To present this test, we need the following definitions. Given training sequences $\bx^N$, a test sequence $y^n$, let
\begin{align}
i^*(\bx^N,y^n)
&:=\argmin_{i\in[M]} \mathrm{GJS}(\hatT_{x_i^N},\hatT_{y^n},\alpha)\label{firsti},\\*
\tilh(\bx^N,y^n)
&:=\min_{\substack{i\in[M]:i\neq i^*(\bx^N,y^n)}} \mathrm{GJS}(\hatT_{x_i^N},\hatT_{y^n},\alpha)\label{def:tildeh}.
\end{align}

Now, given any training sequences $\bx^N$ and test sequence $y^n$, with a appropriately chosen threshold $\tilde{\lambda}$,   Unnikrishnan's  test  (abbreviated as Unn) operates as follows:
\begin{align}
\psi_n^{\rm{Unn}}(\bx^N,y^n)
&:=
\left\{
\begin{array}{ll}
\rmH_j&\mathrm{if}~i^*(\bx^N,y^n)=j ,\tilh(\bx^N,y^n)\geq \tilde{\lambda}\\
\rmH_\rmr&\mathrm{if}~\tilh(\bx^N,y^n)<\tilde{\lambda}.
\end{array}
\right.
\label{def:unntest}
\end{align}
Thus, given  $\bP$,   the type-$j$ error and rejection probabilities for Unnikrishnan's  test  are
\begin{align}
\beta_j(\psi_n^{\rm{Unn}}|\bP)
&=\bbP_j\Big\{i^*(\bX^N,Y^n)\neq j,\tilh(\bX^N,Y^n)\geq \tilde{\lambda}\Big\},\\*
\zeta_j(\psi_n^{\rm{Unn}}|\bP)
&=\bbP_j\Big\{\tilh(\bX^N,Y^n)<\tilde{\lambda}\Big\}.
\end{align}

 Similarly to \eqref{def:typejerror} and \eqref{def:typejreject}, for each $j\in[M]$, we define $\tilde{\bbP}_j\{\cdot\}:=\Pr\{\cdot|\rmH_j\}$ where the training sequences $\bX^N$ are generated from  $\tilde{\bP}$.  For each $j\in[M]$ and for all tuples of distributions $\tilde{\bP}\in\calP(\calX)^M$, we can upper bound the type-$j$ error probability as follows:
\begin{align}
 \beta_j(\psi_n^{\rm{Unn}}|\tilde{\bP} ) &=\tilde{\bbP}_j\Big\{i^*(\bX^N,Y^n)\neq j,~\mathrm{GJS}(X_k^N,Y^n,\alpha)\geq \tilde{\lambda},\forall~k\neq i^*(\bX^N,Y^n)\Big\}\\*
&\leq \tilde{\bbP}_j\Big\{\mathrm{GJS}(X_j^N,Y^n,\alpha)\geq \tilde{\lambda}\Big\}\\
&\leq (n(1+\alpha)+1)^{|\calX|}\exp(-n\tilde{\lambda}),\label{ptypejerror}
\end{align} 
where \eqref{ptypejerror} follows similarly as \eqref{upptype2}.

We then upper bound  the type-$j$ rejection probability   with respect to a particular tuple of distributions $\bP$ satisfying the condition in Theorem \ref{second:cm}. In the following, for brevity, we will use $\imath_1(x|i,j)$ (resp.\ $\imath_2(x|i,j)$) to denote $\imath_1(x|P_i,P_j,\alpha)$ (resp.\ $\imath_2(x|P_i,P_j,\alpha)$) in \eqref{def:i}).

In the following, we will first show that with high probability, the minimizer for $\tilh(\bX^N,Y^n,\alpha)$ in~\eqref{def:tildeh} is given by $i^*(j)$ (see \eqref{def:thetaj}) under hypothesis $\rmH_j$ for each $j\in[M]$. For each $j\in[M]$, we have that
\begin{align}
\nn
&\bbP_j\Big\{\mathrm{GJS}(X_j^N,Y^n,\alpha)>\mathrm{GJS}(X_{i^*(j)}^N,Y^n,\alpha)\Big\}\\
&\leq \bbP_j\bigg\{
-\frac{ 1}{n}\Big(\sum_{k\in[N]}\imath_1(x_{i^*(j),k}|i^*(j),j)+\sum_{k\in[n]}\imath_2(y_k|i^*(j),j)\Big)<O\left(\frac{\log n}{n}\right)\bigg\}+2\tau_n\label{usetaylor}\\
&\leq \rmQ\bigg(\left(\mathrm{GJS}(P_{i^*(j)},P_j,\alpha)+O\left(\frac{\log n}{n}\right)\right)\sqrt{\frac{n}{\rmV(P_i,P_j,\alpha)}}\bigg)+\frac{6\rmT(P_i,P_j,\alpha)}{\sqrt{n(\rmV(P_i,P_j,\alpha))^3}}+2\tau_n\label{useberryagaina},\\
&\leq  \exp\bigg(-\frac{n(\mathrm{GJS}(P_{i^*(j)},P_j,\alpha)+O(\frac{\log n}{n}))^2}{2\rmV(P_{i^*(j)},P_j,\alpha)}\bigg)+\frac{6\rmT(P_{i^*(j)},P_j,\alpha)}{\sqrt{n(\rmV(P_{i^*(j)},P_j,\alpha))^3}}+2\tau_n\label{useineqrmq}\\*
&=:\mu_{1,n}(j)=O\left(\frac{1}{\sqrt{n}}\right)\label{def:mu1n}.
\end{align}
where \eqref{usetaylor} follows similarly to \eqref{usetaylor&atypical} and the fact that $\imath_l(x|j,j)=0$ for $l\in[2]$; \eqref{useberryagaina} follows from the Berry-Esseen theorem similarly to \eqref{useberry} and $\tau_n$ is defined in \eqref{pofatypical}; \eqref{useineqrmq} follows since $\rmQ(x)\leq  \exp (-\frac{x^2}{2})$ for $x\geq 0$; and \eqref{def:mu1n} follows since $\mathrm{GJS}(P_{i^*(j)},P_j,\alpha)> 0$  according to the assumption in Theorem \ref{second:cm} and thus the second term in \eqref{useineqrmq} dominates.

Given any triple of distributions $(P_1,P_2,P_3)\in\calP^3$, let
\begin{align}
\tilde{\rmV}(P_1,P_2,P_3,\alpha)
&:=\alpha\mathrm{Var}_{P_1}[\imath_1(X|1,3)]+\alpha\mathrm{Var}_{P_2}[\imath_1(X|2,3)]+\mathrm{Var}_{P_3}\left[\imath_2(X|1,3)-\imath_2(X|2,3)\right]\label{def:tildermv},\\
\tilde{\rmT}(P_1,P_2,P_3,\alpha)
\nn&:=\alpha\mathbb{E}_{P_1}[|\imath_1(X|1,3)-\mathbb{E}_{P_1}[\imath_1(X|1,3)]|^3]+\alpha\mathbb{E}_{P_2}[|\imath_1(X|2,3)-\mathbb{E}_{P_2}[\imath_1(X|2,3)]|^3]\\*
&\qquad+\mathrm{Var}_{P_3}\left[|\imath_2(X|1,3)-\imath_2(X|2,3)-\mathbb{E}_{P_3}[\imath_2(X|1,3)]+\mathbb{E}_{P_3}[\imath_2(X|2,3)]|^3\right]\label{def:tildermt}.
\end{align}
Similarly to \eqref{def:mu1n}, we have that for each $j\in[M]$ and any $i\in[M]$ s.t. $i\neq j$ and $i\neq i^*(j)$, we have
\begin{align}
\nn&\bbP_j\Big\{\mathrm{GJS}(X_i^N,Y^n,\alpha)<\mathrm{GJS}(X_{i^*(j)}^N,Y^n,\alpha)\Big\}\\*
&\leq \bbP_j\bigg\{
\frac{1}{n}\Big(\sum_{k\in[N]}\big(\imath_1(x_{i^*(j),k}|i^*(j),j)-\imath_1(x_{i,k}|i,j)\big)+\sum_{k\in[n]}\big(\imath_2(y_k|i^*(j),j)-\imath_2(y_k|i,j)\big)\Big)>O\left(\frac{\log n}{n}\right)\bigg\} \nn\\*
&\qquad+2\tau_n\\
\nn&\leq \rmQ\Bigg(\left(\mathrm{GJS}(P_i,P_j,\alpha)-\mathrm{GJS}(P_{i^*(j)},P_j,\alpha)+O\left(\frac{\log n}{n}\right)\right)\sqrt{\frac{n}{\tilde{\rmV}(P_{i^*(j)},P_i,P_j,\alpha)}}\Bigg)\\*
&\qquad+\frac{6\tilde{\rmT}(P_{i^*(j)},P_i,P_j|\alpha)}{\sqrt{n(\tilde{\rmV}(P_{i^*(j)},P_i,P_j,\alpha))^3}} +2\tau_n\label{useberryagainb}\\
&\leq  \exp\Bigg(\! -\! \frac{n\big(\mathrm{GJS}(P_i,P_j,\alpha)\! -\! \mathrm{GJS}(P_{i^*(j)},P_j,\alpha)\! +\! O (\frac{\log n}{n})\big)^2}{2\tilde{\rmV}(P_{i^*(j)},P_i,P_j,\alpha)}\Bigg)+\frac{6\tilde{\rmT}(P_{i^*(j)},P_i,P_j,\alpha)}{\sqrt{n(\tilde{\rmV}(P_{i^*(j)},P_i,P_j,\alpha))^3}} + 2\tau_n\label{beforemu2n}\\
&=:\mu_{2,n}(i,j)=O\left(\frac{1}{\sqrt{n}}\right)\label{def:mu2n},
\end{align}
where \eqref{def:mu2n} holds since $\mathrm{GJS}(P_i,P_j,\alpha)>\mathrm{GJS}(P_{i^*(j)},P_j,\alpha)$ according to assumption that the minimizer for $\theta_j$ (see \eqref{def:thetaj}) is unique and thus the second term in \eqref{beforemu2n} dominates.

For each $j\in[M]$, let
\begin{align}
\mu_n(j)&:=\mu_{1,n}(j)+\sum_{i\in[M]:i\neq j,i\neq i^*(j)}\mu_{2,n}(i,j)=O\left(\frac{1}{\sqrt{n}}\right)\label{def:mun}.
\end{align}
Combining \eqref{def:mu1n} and \eqref{def:mu2n}, we conclude that for each $j\in[M]$,
\begin{align}
\bbP_j\Big\{\tilh(\bX^N,Y^n,\alpha)=\mathrm{GJS}(\hatT_{X_{i^*(j)}^N},\hatT_{Y^n},\alpha)\Big\}\geq 1-\mu_n(j).
\end{align}
Therefore, we have that for each $j\in[M]$,
\begin{align}
 \zeta_j(\psi_n^{\rm{Unn}}|\bP)
&=\bbP_j\Big\{\tilh(\bX^N,Y^n,\alpha)<\tilde{\lambda}\Big\}\\
&\leq \bbP_j\Big\{\mathrm{GJS}(\hatT_{X_{i^*(j)}^N},\hatT_{Y^n},\alpha)<\tilde{\lambda}\Big\}+\mu_n(j)\\
&\leq \Phi\Bigg(\left(\tilde{\lambda}-\mathrm{GJS}(P_{i^*(j)},P_j,\alpha)+O\left(\frac{\log n}{n}\right)\right)\sqrt{\frac{n}{\rmV(P_{i^*(j)},P_j,\alpha)}}\Bigg) \nn\\*
&\qquad+\frac{6\rmT(P_{i^*(j)},P_j,\alpha)}{\sqrt{n(\rmV(P_{i^*(j)},P_j,\alpha))^3}}+\tau_n+\mu_n(j)\label{useberryagain},
\end{align}
where \eqref{useberryagain} follows similarly to \eqref{useberry}, \eqref{useberryagaina} and \eqref{useberryagainb}.

For each $j\in[M]$, let
\begin{align}
\rho_{j,n}&:=\frac{6\rmT(P_{i^*(j)},P_j,\alpha)}{\sqrt{n(\rmV(P_{i^*(j)},P_j,\alpha))^3}}+\tau_n+\mu_n(j)\label{def:rhojn}
\end{align}
Choose $\tilde{\lambda}$ such that
\begin{align}
\tilde{\lambda}&:=
\min_{j\in[M]}\bigg\{\mathrm{GJS}(P_{i^*(j)},P_j,\alpha)+\sqrt{\frac{\rmV(P_{i^*(j)},P_j,\alpha)}{n}}\Phi^{-1}(\varepsilon_j-\rho_{j,n})\bigg\}+O\left(\frac{\log n}{n}\right)\label{choose:tlj},
\end{align}
and let
\begin{align}
\lambda&:=\tilde{\lambda}-\frac{|\calX|\log (n(1+\alpha)+1)}{n}\label{def:choose:lj}.
\end{align}
Invoking the results in \eqref{ptypejerror}, \eqref{useberryagain} and applying a Taylor expansions to $\Phi^{-1}(\cdot)$ (similarly to \eqref{taylorphi}), we conclude that Unnikrishnan's  test  $\psi_n^{\rm{Unn}}$ in \eqref{def:unntest} satisfies the following two conditions:
\begin{itemize}
\item for all tuples of distributions $\tilde{\bP}\in\calP(\calX)^M$ and for each $j\in[M]$, $\beta_j(\psi_n|\tilde{\bP})\leq \exp(-n\lambda)$ ;
\item for any tuple of distributions $\bP$ satisfying the condition in Theorem \ref{second:cm}, $\zeta_j(\psi_n|\bP)\leq \varepsilon_j$.
\end{itemize}
The achievability proof of Theorem \ref{second:cm} is completed.

\subsubsection{Converse Proof}
Given any $\bm{\kappa}=(\kappa_1,\ldots,\kappa_M)\in[0,1]^M$, let 
\begin{align}
\underline{\kappa}=\min_{t\in[M]}\kappa_t,\quad\mbox{and}\quad \kappa_{+}=\sum_{t\in[M]}\kappa_t.
\end{align}
Paralleling Lemma \ref{anytotype}, we relate the error and rejection probabilities of any arbitrary test to a type-based test (i.e., the test is a  function of only the marginal types  $(\hatT_{X_1^N},\ldots,\hatT_{X_M^N},\hatT_{Y^n})$). 
\begin{lemma}
\label{anytotype:cm}
Given any arbitrary test $\psi_n$ and any $\bm{\kappa}\in[0,1]^M$, for any tuple of distributions $\bP\in\calP(\calX)^M$, we can construct a type-based test $\psi_n^\rmT$ such that for each $j\in[M]$,
\begin{align}
\beta_j(\psi_n|\bP)&\geq \underline{\kappa}\beta_j(\psi_n^\rmT|\bP),\\
\zeta_j(\psi_n|\bP)&\geq (1-\kappa_{+})\zeta_j(\psi_n^\rmT|\bP).
\end{align}
\end{lemma}
The proof of Lemma \ref{anytotype:cm} is analogous to that of Lemma~\ref{anytotype} and is thus omitted.

Paralleling Lemma \ref{typeconverse}, in the following lemma, we derive a lower bound on type-$j$ rejection probability for each $j\in[M]$ with respect to a particular tuple of distributions for any type-based test satisfying that type-$j$ error probability decays exponentially fast for each $j\in[M]$ and for all tuples of distributions.

Recall the definition of $\tilh(\cdot)$ in \eqref{def:tildeh}. For simplicity, let
\begin{align}
\eta_{n,M}&:=\frac{M|\calX|\log (n\alpha+1)}{n\alpha}+\frac{|\calX|\log(n+1)}{n}\label{def:etanM}.
\end{align}
\begin{lemma}
\label{typeconverse:cm}
For any $\lambda\in\bbR_+$ and any type-based test $\psi_n^\rmT$ such that for all tuples of distributions $\tilde{\bP}\in\calP(\calX)^M$, 
\begin{align}
\beta_j(\psi_n^\rmT|\tilde{\bP})\leq \exp(-n\lambda),\quad \forall \, j\in[M]\label{type:constraint:cm},
\end{align}
we have that for any particular tuple of distributions $\bP\in\calP(\calX)^M$,
\begin{align}
\zeta_j(\psi_n^\rmT)
&\geq \bbP_j\Big\{\tilh(\bX^N,Y^n)+\eta_{n,M}<\lambda\Big\},\quad\forall\, j\in[M].
\end{align}
\end{lemma}
The proof of Lemma \ref{typeconverse:cm} is similar to that for Lemma \ref{typeconverse} and so it is omitted.

Combining  Lemmas \ref{anytotype:cm} and \ref{typeconverse:cm} and letting $\kappa_j=1/n$ for each $j\in[M]$, for any test $\psi_n$ satisfying that for all $\tilde{\bP}\in\calP(\calX)^M$,
\begin{align}
\beta_j(\psi_n|\tilde{\bP})\leq \exp(-n\lambda),\quad\forall\, j\in[M],\label{converse:cm}
\end{align}
given any tuple of distributions $\bP\in\calP(\calX)^M$, we have that for each $j\in[M]$,
\begin{align}
\zeta_j(\psi_n|\bP)&\geq \left(1-\frac{M}{n}\right) \bbP_j\Big\{\tilh(\bX^N,Y^n)+\eta_{n,M}+\frac{\log n}{n}<\lambda\Big\}.
\end{align}
The rest of the converse proof for Theorem \ref{second:cm} is completed  similarly to the achievability part.

\subsection{Proof of Proposition \ref{dual:unn}}
\label{proof:dual:unn}
The proof of Proposition \ref{dual:unn} is similar to that of Proposition \ref{weakc4bc}. Recall Gutman's test in \eqref{gut:mwithreject} and the notations in Section \ref{proof:weakc4bc}. 
 Given any triple of distributions $(P_j,P_i,P_k)\in\calP(\calX)^3$ and any $\alpha\in\bbR_+$, define
\begin{align}
K(P_j,P_i,P_k,\lambda)
&:=\min_{\substack{(Q_1,Q_2,Q_3)\in\calP(\calX)^3:\\\mathrm{GJS}(Q_2,Q_1,\alpha)\leq \lambda\\\mathrm{GJS}(Q_3,Q_1,\alpha)\leq \lambda}} \big\{D(Q_1\|P_j)+\alpha D(Q_2\|P_i)+\alpha D(Q_3\|P_k)\big\}.
\end{align}
Using the KKT conditions~\cite{boyd2004convex} and the definition of $D_{\gamma}(\cdot,\cdot,\cdot)$  in \eqref{def:gdivergence}, one can easily verify that
\begin{align}
K(P_j,P_i,P_k,0)
&=\min_{Q\in\calP(\calX)} \big\{D(Q\|P_j)+\alpha D(Q\|P_i)+\alpha D(Q\|P_k)\big\}=D_{\frac{2\alpha}{1+2\alpha}}(P_j,P_i,P_k)\label{imtouse}.
\end{align}

\subsubsection{Achievability Proof}
In the following analysis, we choose $\lambda$ as in \eqref{lambddainwca}. For any $j\in[M]$ and any $\tilde{\bP}\in\calP(\calX)^M$, given any $\varepsilon\in(0,1)$, we can upper bound $j$-th error probability as follows:
\begin{align}
\limsup_{n\to\infty}\beta_j(\Psi_n^{\rm{Gut}}|\tilde{\bP})
&\leq \limsup_{n\to\infty} \bbP_j \Big\{\mathrm{GJS}(\hatT_{X_j^N},\hatT_{Y^n},\alpha)>\lambda\Big\}\\*
&=\limsup_{n\to\infty} \Pr\Big\{Z>\rmG_{|\calX|-1}^{-1}(\varepsilon)\Big\}=\varepsilon\label{useweakcga},
\end{align}
where \eqref{useweakcga} follows from the weak convergence analysis in Unnikrishnan and Huang~\cite{unnikrishnan2016weak}.

Furthermore, for any $j\in[M]$ and for any $\bP\in\calP(\calX)^M$, the $j$-th rejection probability satisfies that
\begin{align}
\zeta_j(\Psi_n^{\rm{Gut}}|\bP)
&=\bbP_j\Big\{\exists~(i,k)\in\calM\mathrm{~s.t.~}\mathrm{GJS}(\hatT_{X_i^N},\hatT_{Y^n},\alpha)\leq\lambda,~\mathrm{GJS}(\hatT_{X_k^n},\hatT_Y^n,\alpha)\leq\lambda\Big\}\\
&\leq \frac{M(M-1)}{2}\max_{(j,k)\in\calM}\bbP_j\Big\{\mathrm{GJS}(\hatT_{X_i^N},\hatT_{Y^n},\alpha)\leq\lambda,~\mathrm{GJS}(\hatT_{X_k^n},\hatT_Y^n,\alpha)\leq\lambda\Big\}\\
&=\frac{M(M-1)}{2}(N+1)^{2|\calX|}(n+1)^{|\calX|}\exp\Big(-n\min_{(i,k)\in\calM} K(P_j,P_i,P_k,\lambda)\Big),\label{methodoftypes}
\end{align}
where \eqref{methodoftypes} follows similarly to \eqref{methodoftypeseasy}. Hence, using the choice of $\lambda$ in \eqref{lambddainwca}, the equality in \eqref{imtouse}, and the continuity of $\lambda\mapsto K(P_j,P_i,P_k,\lambda)$ at $\lambda=0$, we have that for each $j\in[M]$,
\begin{align}
\liminf_{n\to\infty}-\frac{1}{n}\log \zeta_j(\Psi_n^{\rm{Gut}}|\bP)
&\geq \min_{(i,k)\in\calM}K(P_j,P_i,P_k,0)=\min_{(i,k)\in\calM} D_{\frac{2\alpha}{1+2\alpha}}(P_j,P_i,P_k).\label{eqn:ach_M}
\end{align}

\subsubsection{Converse Proof for Gutman's Test}
For any $j\in[2:M]$, given any $\tilde{\bP}\in\calP(\calX)^M$, using the union bound, we can lower bound the $j$-th error probability as  follows:
\begin{align}
\beta_j(\Psi_n^{\rm{Gut}}|\tilde{\bP})
&\geq \tilde{\bbP}_j\Big\{\Psi_n^{\rm{Gut}}(\bX^M,Y^n)=\rmH_1\Big\}\\
&\geq \tilde{\bbP}_j\Big\{\mathrm{GJS}(\hatT_{X_i^N},\hatT_{Y^n},\alpha)> \lambda,~\forall~i\in[2:M]\Big\}\\
&\geq \tilde{\bbP}_j\Big\{\mathrm{GJS}(\hatT_{X_j^N},\hatT_{Y^n},\alpha)>\lambda\Big\}-\sum_{i\in[2:M] \setminus\{ j\}}\tilde{\bbP}_j\Big\{\mathrm{GJS}(\hatT_{X_i^N},\hatT_{Y^n},\alpha)\leq\lambda\Big\}. \label{eqn:lwb}
\end{align}
Assume  that $\lambda$ satisfies
\begin{equation}
\lambda=\frac{1}{2n}\rmG_{|\calX|-1}^{-1}(\varepsilon+\delta)
\end{equation}
for some $\delta>0$.  Compare this choice to~\eqref{lambddainwca}. Since $\mathrm{GJS}(\hatT_{X_i^N},\hatT_{Y^n},\alpha)$ converges in probability to $\mathrm{GJS}(\tilP_i, \tilP_j, \alpha)>0$ under $\tilde{\bbP}_j$ and $\lambda\downarrow 0$, all the terms in the sum in \eqref{eqn:lwb} vanish. In addition, by weak convergence (cf.~\eqref{weakconjh}), the first term in \eqref{eqn:lwb} converges to $\varepsilon+\delta>\varepsilon$, contradicting the requirement that $\max_{j\in[M]}\beta_j(\Psi_n^{\rm{Gut}}|\tilde{\bP})\leq \varepsilon$ for all $\tilde{\bP}$; see~\eqref{def:caltau2}. Therefore,  to fulfil this requirement, the threshold $\lambda$ in Gutman's test in~\eqref{gut:mwithreject}  must satisfy
\begin{align}
\lambda\ge \frac{1}{2n}\rmG_{|\calX|-1}^{-1}(\varepsilon),\label{eqn:choice_lam}
\end{align}
because $\rmG_{|\calX|-1}^{-1}(\cdot)$ is monotonically non-increasing (cf.~Section~\ref{sec:notation}).
Furthermore, for each $j\in[M]$, we can lower bound the $j$-th rejection probability as follows:
\begin{align}
\zeta_j(\Psi_n^{\rm{Gut}}|\bP)
&\geq \max_{(i,k)\in\calM}\bbP_j\Big\{\mathrm{GJS}(\hatT_{X_i^N},\hatT_{Y^n},\alpha)\leq\lambda,~\mathrm{GJS}(\hatT_{X_k^n},\hatT_Y^n,\alpha)\leq\lambda\Big\}\\*
&\geq \max_{(i,k)\in\calM}\exp \Big(-nD_{\frac{2\alpha}{1+2\alpha}}(P_j,P_i,P_k)+\Theta(\log n)\Big)\label{similartech},
\end{align}
where \eqref{similartech} follows from~\eqref{eqn:choice_lam} and steps similar to those that led to~\eqref{decreasinl} and Lemma \ref{fn<=f+}. Hence, for each $j\in[M]$, 
\begin{align}
\limsup_{n\to\infty}-\frac{1}{n}\log\max_{j\in[M]}\zeta_j(\Psi_n^{\rm{Gut}}|\bP)
\leq \min_{(i,k)\in\calM}D_{\frac{2\alpha}{1+2\alpha}}(P_j,P_i,P_k).
\end{align}
This and \eqref{eqn:ach_M}  complete the proof of Proposition~\ref{dual:unn}.

\appendix
\subsection{Proof of Lemma \ref{anytotype}}
\label{proof:anytotype}

In the following, for simplicity, we let $\bQ=(Q_1,Q_2,Q_3)\in\calP_N^2(\calX)\times\calP_n(\calX)$. Furthermore, for any $\bQ$, we use $\calT^{n+2N}_{\bQ}$ to denote the set of sequence triples $(x_1^N,x_2^N,y^n)$ such that $x_1^N\in\calT^N_{Q_1}$, $x_2^N\in\calT^N_{Q_2}$ and $y^n\in\calT^n_{Q_3}$. For any test $\phi_n$, we can construct a type-based test $\phi_n^\rmT$ as follows.

Given any type triple $\bQ\in\calP_N^2(\calX)\times\calP_n(\calX)$, if at least $\kappa$ fraction of the sequence triples in the type class $\calT^{n+2N}_Q$ are in the rejection region of the test $\phi_n$, i.e., $|\calT^{n+2N}_{\bQ}\cap\calA^{\rmc}(\phi_n)|\geq \kappa|\calT^{n+2N}_{\bQ}|$, then we let $\phi_n^\rmT(\bQ)=\rmH_2$; otherwise, $\phi_n^\rmT(\bQ)=\rmH_1$.

For any pair of distributions $(P_1,P_2)\in\calP(\calX)^2$, we can then relate the error probabilities of the test $\phi_n^\rmT$ and the original test $\phi_n$ as follows:
\begin{align}
\beta_1(\phi_n|P_1,P_2)
&=\bbP_1\Big\{\phi_n(Y^n,X_1^N,X_2^N)=\rmH_2\Big\}\\*
&=\sum_{\bQ }\bbP_1\Big\{\calA^\rmc(\phi_n)\cap\calT^{n+2N}_{\bQ}\Big\}\\
&\geq \sum_{\bQ: |\calT^{n+2N}_{\bQ}\cap\calA^{\rmc}(\phi_n)|\geq \kappa|\calT^{n+2N}_{\bQ}|}\bbP_1\Big\{\calA^\rmc(\phi_n)\cap\calT^{n+2N}_{\bQ}\Big\}\\
&\geq \sum_{ \bQ : |\calT^{n+2N}_{\bQ}\cap\calA^{\rmc}(\phi_n)|\geq \kappa|\calT^{n+2N}_{\bQ}|}\kappa \bbP_1\Big\{\calT^{n+2N}_{\bQ}\Big\}\\
&\geq \kappa \beta_1(\phi_n^\rmT|P_1,P_2) 
\end{align}
and
\begin{align}
\beta_2(\phi_n|P_1,P_2)
&=\bbP_2\Big\{\calA(\phi_n)\Big\}\\*
&\geq \sum_{\bQ:   |\calT^{n+2N}_{\bQ}\cap\calA^{\rmc}(\phi_n)|<\kappa|\calT^{n+2N}_{\bQ}|}\bbP_2\Big\{\calA(\phi_n)\cap\calT_{\bQ}\Big\}\\
&\geq \sum_{ \bQ  : |\calT^{n+2N}_{\bQ}\cap\calA^{\rmc}(\phi_n)|<\kappa|\calT^{n+2N}_{\bQ}|} (1-\kappa)\bbP_2\Big\{\calT^{n+2N}_{\bQ}\Big\}\\
&=(1-\kappa)\beta_2(\phi_n^\rmT).
\end{align}
This completes the proof of Lemma \ref{anytotype}.

\subsection{Proof of Lemma \ref{typeconverse}}
\label{proof:typeconverse}
We claim that for any type-based test $\phi_n^\rmT$ satisfying \eqref{typeconstraint}, if a type triple $\bQ$ satisfies
\begin{align}
\mathrm{GJS}(Q_1,Q_3,\alpha)+\frac{|\calX|\log (n+1)}{n}+\frac{2|\calX|\log(1+\alpha n)}{\alpha n}<\lambda\label{accept:Q},
\end{align}
then we have $\phi_n^\rmT(\bQ)=\rmH_1$.

This can be proved by contradiction. Suppose our claim in \eqref{accept:Q} {\em were} not true, then there exists a type triple $\bar{\bQ}=(\barQ_1,\barQ_2,\barQ_3)$ such that 
\begin{align}
\mathrm{GJS}(\barQ_1,\barQ_3,\alpha)+\frac{|\calX|\log (n+1)}{n}+\frac{2|\calX|\log(1+\alpha n)}{\alpha n}<\lambda,\quad\mbox{and}\quad\phi_n^\rmT(\bar{\bQ})=\rmH_2.
\end{align}
Therefore, we have that for all $(\tilP_1,\tilP_2)\in\calP(\calX)^2$,
\begin{align}
\beta_1(\phi_n^\rmT|\tilP_1,\tilP_2)
&=\tilde{\bbP}_1\Big\{\phi_n^{\rmT}(\hatT_{X_1^N},\hatT_{X_2^N},\hatT_{Y^n})=\rmH_2\Big\}\\
&=\sum_{\bQ :\phi_n^\rmT(\bQ)=\rmH_2}\tilP_1^n(\calT^n_{Q_1})\tilP_2^N(\calT^N_{Q_2})\tilP_1^N(\calT^N_{Q_3})\\
&\geq \tilP_1^n(\calT^N_{\barQ_1})\tilP_2^N(\calT^N_{\barQ_2})\tilP_1^n(\calT^n_{\barQ_3})\\
&\geq (n+1)^{-|\calX|}(N+1)^{-2|\calX|}\exp\Big\{-ND(\barQ_1\|\tilP_1)-ND(\barQ_2\|\tilP_2)-nD(\barQ_3\|\tilP_1)\Big\}.
\end{align}
However, if we let $\tilP_1= \frac{1}{1+\alpha}(\alpha\barQ_1+\barQ_3)$ and $\tilP_2=\barQ_2$, then
\begin{align}
\beta_1(\phi_n^\rmT|\tilP_1,\tilP_2)
&\geq (n+1)^{-|\calX|}(N+1)^{-2|\calX|}\exp(-n\mathrm{GJS}(\barQ_1,\barQ_3,\alpha))\\
&=\exp\bigg(-n \bigg(\mathrm{GJS}(\barQ_1,\barQ_3,\alpha)+\frac{|\calX|\log (n+1)}{n}+\frac{2|\calX|\log(1+\alpha n)}{\alpha n}\bigg)\bigg)\\
&>\exp(-n\lambda),
\end{align}
which contradicts the assumption that~\eqref{typeconstraint} holds for any $(P_1,P_2)\in\calP(\calX)^2$. Thus, we have shown that for any  $(Q_1,Q_2,Q_3)$ satisfying \eqref{accept:Q}, given any type-based test $\phi_n^\rmT$ satisfying \eqref{typeconstraint} for all $(\tilP_1,\tilP_2)\in\calP(\calX)$, we have $\phi_n^\rmT(Q_1,Q_2,Q_3)=\rmH_1$.

\subsection{Proof of Lemma \ref{fn<=f+}}
\label{proof:fn<=f+}
Recall that $Q^*=P^{(\frac{\alpha}{1+\alpha})}$ achieves $F(P_1,P_2,\lambda,0)=D_{\frac{\alpha}{1+\alpha}}(P_1\|P_2)$. Also recall that $n':=\min\{n,N\}$. We can find a type $\barQ\in\calP_{n'}(\calX)$ such that for any $x\in\calX$, $|\barQ(x)-Q^*(x)|\leq {1}/{n'}$. Then using the definition of $F_n(P_1,P_2,\alpha,\lambda)$ in \eqref{def:Fn} we have
\begin{align}
&F_n(P_1,P_2,\alpha,0)\nn\\*
&\leq \alpha D(\barQ\|P_1)+D(\barQ\|P_2)\\
&\leq \alpha D(Q^*\|P_1)+ D(Q^*\|P_2)+\alpha |D(Q^*\|P_1)-D(\barQ\|P_1)|+|D(Q^*\|P_2)-D(\barQ\|P_2)|\\
&\leq D_{\frac{\alpha}{1+\alpha}}(P_1\|P_2)+\frac{(1+\alpha)|\calX|}{n'}\log n'-\frac{\sum_x \log (P_1^\alpha(x)P_2(x))}{n'}\label{uselemma2.7}
\end{align}
where \eqref{uselemma2.7} follows from \cite[Lemma 1.2.7]{csiszar2011information} and the fact that $\sum_x |\barQ(x)-Q^*(x)|\leq  {|\calX|}/{n'}$.

\section*{Funding}
This work was funded by an National University of Singapore (NUS) Grant  (C-261-000-207-532 and C-261-000-005-001)  and a Singapore National Research Foundation (NRF) Fellowship (NRF2017NRF-NRFF001-070 and R-263-000-D02-281).

\bibliographystyle{IEEEtran}
\bibliography{IEEEfull_lin}
\end{document}